\documentclass[11pt,preprint]{aastex}

\begin{document}

\shorttitle{Disk Structures in Ophiuchus}
\shortauthors{Andrews et al.}

\title{Protoplanetary Disk Structures in Ophiuchus}

\author{Sean M. Andrews\altaffilmark{1,2}, D. J. Wilner\altaffilmark{1}, A. M. Hughes\altaffilmark{1}, Chunhua Qi\altaffilmark{1}, and C. P. Dullemond\altaffilmark{3}}

\altaffiltext{1}{Harvard-Smithsonian Center for Astrophysics, 60 Garden Street, Cambridge, MA 02138; sandrews, dwilner, mhughes, cqi@cfa.harvard.edu}
\altaffiltext{2}{Hubble Fellow}
\altaffiltext{3}{Max Planck Institut f\"{u}r Astronomie, K\"{o}nigstuhl 17, 69117 Heidelberg, Germany; dullemon@mpia.de}

\begin{abstract}
We present the results of a high angular resolution (0\farcs3 $\approx$ 40\,AU) 
Submillimeter Array survey of the 345\,GHz (870\,$\mu$m) thermal continuum 
emission from 9 of the brightest, and therefore most massive, circumstellar 
disks in the $\sim$1\,Myr-old Ophiuchus star-forming region.  Using 
two-dimensional radiative transfer calculations, we simultaneously fit the 
observed continuum visibilities and broadband spectral energy distribution for 
each disk with a parametric structure model.  Compared to previous millimeter 
studies, this survey includes significant upgrades in modeling, data quality, 
and angular resolution that provide improved constraints on key structure 
parameters, particularly those that characterize the spatial distribution of 
mass in the disks.  In the context of a surface density profile motivated by 
similarity solutions for viscous accretion disks, 
$\Sigma\propto(R/R_c)^{-\gamma}\exp{[-(R/R_c)^{2-\gamma}]}$, the best-fit 
models for the sample disks have characteristic radii $R_c \approx 20$-200\,AU, 
high disk masses $M_d \approx 0.005$-0.14\,M$_{\odot}$ (a sample selection 
bias), and a narrow range of radial $\Sigma$ gradients ($\gamma \approx 
0.4$-1.0) around a median $\gamma = 0.9$.  These density structures are used in 
conjunction with accretion rate estimates from the literature to help 
characterize the viscous evolution of the disk material.  Using the standard 
prescription for disk viscosities, those combined constraints indicate that 
$\alpha \approx 0.0005$-0.08.  Three of the sample disks show large ($R \approx 
20$-40\,AU) central cavities in their continuum emission morphologies, marking 
extensive zones where dust has been physically removed and/or has significantly 
diminished opacities.  Based on the current requirements of planet formation 
models, these emission cavities and the structure constraints for the sample as 
a whole suggest that these young disks may eventually produce planetary 
systems, and have perhaps already started.  
\end{abstract}
\keywords{accretion, accretion disks --- circumstellar matter --- planetary 
systems: protoplanetary disks --- solar system: formation --- stars: 
pre-main-sequence}

\section{Introduction}

Within a span of $\sim$10\,Myr, the material in a circumstellar disk will 
either be accreted onto its star, dispersed into the interstellar medium, or 
incorporated into the larger bodies of a burgeoning planetary system.  
Throughout that time, viscous and gravitational forces spatially redistribute 
this material and its angular momentum, driving a net inward mass flow that can 
accrete onto the stellar surface \citep[e.g.,][]{pringle81}.  Meanwhile, dust 
grains settle to the disk midplane and accumulate into larger solid bodies
\citep{beckwith00,dullemond05}.  If that growth progresses sufficiently, those 
bodies can dynamically influence the remaining disk material through accretion, 
ejection, and the sculpting of gaps and large-scale cavities 
\citep{lin93,papaloizou07}.  Toward the end of the disk lifetime, winds driven 
by energetic radiation from the central star can rapidly sweep any remaining 
material into the local environment \citep{clarke01,alexander06a,alexander07}.  
Each of these evolutionary mechanisms is critically influenced by the spatial 
mass distribution $-$ the density structure $-$ of the disk material.

While the theoretical machinery behind these evolution processes has been 
developed in some detail \citep[see][]{dullemond07,youdin08,alexander08}, 
empirical constraints on densities in circumstellar disks are still notably 
rare.  Observational limitations are largely to blame, in part due to the small 
angles that disks subtend on the sky and the inability to directly observe the 
dominant mass constituent (molecular hydrogen).  In light of these 
deficiencies, many studies use a crude approximation of the primordial solar 
disk (the Minimum Mass Solar Nebula, or MMSN) as a reference point.  A surface 
density profile ($\Sigma$) for the MMSN is constructed by augmenting the 
current planet masses to match solar abundances, and then smearing those masses 
into concentric annuli.  The result is usually fit with a radial power-law, 
$\Sigma \propto R^{-3/2}$ \citep[e.g.,][]{weidenschilling77,hayashi81}.  
Spatially resolved observations of the dust in disks around nearby young stars 
can provide more direct constraints on a potentially wide variety of density 
structures.  Measurements at millimeter wavelengths are particularly desirable, 
as the thermal continuum emission from dust grains is optically thin and 
therefore probes the full disk volume \citep{beckwith90,beckwith91}.  For a 
simple prescription of that emission, $S_{\nu} \propto B_{\nu} (1-e^{-\tau})$, 
where $S_{\nu}$ is the surface brightness, $B_{\nu}$ the Planck function at the 
local temperature ($T$), and $\tau$ the optical depth.  The low millimeter 
optical depths simplify the reprocessing term ($1-e^{-\tau} \approx \tau$) and 
imply that the emission is produced near the cold disk midplane.  Assuming the 
Rayleigh-Jeans approximation is valid ($B_{\nu} \propto T$) and defining the 
optical depth as the product of the dust opacity and column (surface) density 
($\tau = \kappa \Sigma$), we note that the millimeter continuum emission tracks 
a compound product of physical conditions in the disk, $S_{\nu} \propto \kappa 
\Sigma T$.  

Although associating the emission and disk structure in this way is simplistic, 
it has served as the intuitive basis for interpreting a wealth of millimeter 
observations.  With some disk-averaged assumptions for $\kappa$ and $T$, it has 
been used to estimate hundreds of disk masses from single-dish millimeter 
photometry surveys \citep{beckwith90,andre94,osterloh95,aw05,aw07b}.  
Interferometric studies have typically taken a more sophisticated approach and 
fitted resolved millimeter data with parametric disk structure models, where 
$\Sigma \propto R^{-p}$.  A wide variety of surface density profiles, with $p$ 
values ranging from 0 to 2, have been derived from such fits, based on the 
millimeter continuum data alone \citep{lay94,lay97,mundy96} or in combination 
with the broadband spectral energy distribution 
\citep[SED;][]{wilner00,testi01,akeson02,kitamura02,aw07}.  Similar results 
have been determined from resolved spectral images of line emission from trace 
molecular species, which suffer from the additional complexities of excitation 
and abundance variations \citep{dutrey98,dutrey03,guilloteau98,guilloteau99,simon00,dartois03,pietu03,pietu07,isella07}.  

While those studies have profoundly shaped our knowledge of disk structure, 
they are all fundamentally limited by low angular resolution.  Typical 
interferometer baselines probed angular scales down to 1-4\arcsec, 
corresponding to 125-500\,AU for the nearest star-forming regions.  With disk 
diameters in the 100-1000\,AU range, these previous observations simply did not 
have the spatial dynamic range required to accurately characterize disk 
emission morphologies, and therefore disk structures.  For such limited angular 
resolution, the emission could only be probed at large disk radii ($R \ge 
60$\,AU at best), well outside the region where the density structure is most 
relevant to the viscous evolution and planet formation processes ($R \le 
40$\,AU).  Some recent studies have overcome these restrictions with long 
interferometer baselines and provided the first millimeter-wave views of 
structure in the inner disk ($R \approx 20$-50\,AU) with sub-arcsecond 
resolution images 
\citep[0.3-0.5\arcsec;][]{hamidouche06,pietu06,guilloteau08,andrews08}.  A few 
such observations have even discovered large central cavities in the dust 
emission $-$ potential signatures of young planetary systems 
\citep{pietu06,brown08,brown09,hughes07,hughes09}.  These novel high angular 
resolution capabilities offer new insights on disk structures that can help 
refine our understanding of both disk evolution and planet formation.

With those goals in mind, we present the results of a new high angular 
resolution (0\farcs3) survey of the 345\,GHz (870\,$\mu$m) continuum emission 
from 9 circumstellar disks in the $\sim$1\,Myr-old Ophiuchus star formation 
region.  We use these data to extract constraints on the disk structures in the 
spirit of the study by \citet{aw07}, but now with major improvements in the 
data quality, spatial resolution, and modeling techniques.  The interferometric 
survey observations and data calibration are described in \S 2.  Our disk model 
calculations are introduced in \S 3, and the resulting constraints on key disk 
structure parameters are presented in \S 4.  The derived disk structures are 
examined in the contexts of their viscous properties and planet formation 
prospects in \S 5.  A summary of our principal conclusions is provided in \S 6.

\section{Observations and Data Reduction}

A sample of 9 disks was observed with the very extended (V) configuration of 
the Submillimeter Array interferometer \citep[SMA;][]{ho04} at Mauna Kea, 
Hawaii.  In this array configuration, the eight 6\,m SMA antennas span 
baselines of 68-509\,m.  Double sideband receivers were tuned to a local 
oscillator (LO) frequency of 340.755\,GHz (880\,$\mu$m).  Each sideband 
contains 24 partially overlapping 104\,MHz chunks centered $\pm$5\,GHz from 
the LO frequency.  Similar observations were also obtained in the sub-compact 
(S), compact (C), and extended (E) SMA configurations, providing baseline 
lengths of 6-70\,m, 16-70\,m, and 28-226\,m, respectively.  Some of the S, C, 
and E observations had slightly different receiver tunings and correlator 
setups in an effort to sample the CO $J$=3$-$2 transition on a finer spectral 
resolution scale.  A journal of the SMA observations is provided in Table 
\ref{obs_journal}.  The C and E data obtained prior to 2007 were already 
described by \citet{aw07}.  The V data for SR 21 in the SMA archive were 
originally presented by \citet{brown09}.

The observing sequence interleaved disk targets and (at least two) quasars in 
an alternating pattern with a 2:1 integration time ratio.  The total cycle time 
between quasar observations in the V configuration was limited to 8 minutes to 
ensure that any short-timescale phase variations could be appropriately 
calibrated.  A longer cycle ($\sim$15-20 minutes) was used for the S, C, and E 
observations.  Additional calibrators were observed when the targets were at 
low elevations ($<$20\degr).  Depending on their proximity to the disk targets 
and fluxes at the time of the observations, we chose from a group of four 
quasars to be used as gain calibrators: J1625$-$254, J1626$-$298, J1517$-$243, 
and J1733$-$130.  Planets (Uranus, Jupiter, Saturn), satellites (Titan, 
Callisto), and bright quasars (3C 454.3, 3C 279) were observed as bandpass and 
absolute flux calibrators depending on their availability and the array 
configuration.  The observing conditions in the V configuration were excellent, 
with atmospheric opacities $\le$0.05 at 225\,GHz (corresponding to $\le$1.0\,mm 
of precipitable water vapor) and well-behaved phase variations on timescales 
longer than the calibration cycle.  

The data were edited and calibrated with the IDL-based {\tt MIR} software package.\footnote{\url{See http://cfa-www.harvard.edu/$\sim$cqi/mircook.html.}}  The 
bandpass response was calibrated with observations of a bright planet or 
quasar, and broadband continuum channels in each sideband were generated by 
averaging the central 82\,MHz in all line-free chunks.  The visibility 
amplitude scale was set based on observations of planets/satellites (Uranus, 
Titan, or Callisto) and routinely-monitored quasars: the typical systematic 
uncertainty in the absolute flux scale is $\sim$10\%.  The antenna-based 
complex gain response of the system as a function of time was determined with 
reference to the quasar nearest on the sky to the corresponding disk target.  
The other quasar in the observing cycle provides a check on the quality of the 
phase transfer in the gain calibration process.  We find that the millimeter 
``seeing" generated by atmospheric phase noise and any small baseline errors is 
minimal, 0\farcs1 at the most.  Because the visibilities for all disk targets 
show excellent agreement between sidebands and on the overlapping baselines for 
different configurations, all data for a given target were combined.  The 
standard tasks of Fourier inverting the visibilities, deconvolution with the 
{\tt CLEAN} algorithm, and restoration with a synthesized beam were conducted 
with the {\tt MIRIAD} software package.  Maps of the continuum emission were 
created with a Briggs robust = 0.7 weighting scheme for the visibilities, and 
maps of the CO $J$=3$-$2 line emission were made with natural weighting (robust 
= 2) for the S, C, and E data when available.  Some of the relevant data 
properties from these synthesized maps are compiled in Table \ref{data_table}.

The synthesized continuum maps for the sample disks are featured together in 
Figure \ref{images}.  Each high angular resolution map covers 4\arcsec\ on a 
side, corresponding to 500\,AU at the adopted distance of 125\,pc to the 
Ophiuchus clouds \citep{degeus89,knude98,lombardi08,loinard08}.  A centroid 
position and initial estimate of the viewing geometry of the disk $-$ 
characterized by the inclination ($i$) and major axis position angle (PA) $-$ 
were determined by fitting the visibilities with an elliptical Gaussian 
brightness distribution.  Because the continuum emission from the DoAr 44 and 
SR 21 disks is obviously not centrally peaked, their centroid positions and 
viewing geometries were estimated by inspection of the images.  In most cases, 
any CO line emission from the disks is significantly contaminated by the local 
molecular cloud environment.  Figure \ref{moments} shows CO $J$=3$-$2 moment 
maps for the 3 disks that suffer the least such contamination.  Note that the 
image scale in those panels, 10\arcsec (1250\,AU) on a side, is significantly 
larger than for the continuum maps in Figure \ref{images}.  The contours 
represent the velocity-integrated intensity (zeroth moment) and the color scale 
the intensity-weighted velocities (first moment).  While CO emission was 
detected in most cases (except for the SR 21 and WSB 60 data), its association 
with the disks rather than the molecular cloud is unclear.

\section{Modeling the Disk Structures}

The high angular resolution images exhibited in Figure \ref{images} reveal a 
striking diversity of continuum emission morphologies, ranging from centrally 
concentrated and compact (e.g., VSSG 1) to more diffuse and extended (e.g., AS 
209) and even cases with large central depressions (SR 21 and DoAr 44).  But 
understanding how the underlying disk structures are related to these 
morphologies requires a more sophisticated interpretation.  Our strategy for 
this task is to simultaneously reproduce the SMA continuum visibilities and 
broadband SED for each disk with a parametric structure model.  While this is 
the same concept used in a previous SMA disk survey by \citet{aw07}, our study 
makes some major improvements in the model calculations.  One critical upgrade 
is the combined use of a two-dimensional structure model for flared disks and a 
Monte Carlo radiative transfer code to compute synthetic observations.  Rather 
than the typical {\it ad hoc} temperature parameterization imposed on such 
models, the radiative transfer code yields a temperature structure that is 
internally consistent with a given parametric density structure.  These 
calculations are especially important when fitting a SED, as they provide a 
proper accounting of emission contributions at each frequency from all of the 
relevant regions in the disk.  In the remainder of this section, we introduce 
the parametric model for the disk density structures (\S 3.1), highlight some 
critical assumptions made in the modeling (\S 3.2), and explain the radiative 
transfer calculations (\S 3.3) and methodology used to estimate model 
parameters (\S 3.4).

\subsection{The Density Structure Model}

The disk structure model and radiative transfer calculations are defined on a 
spatial grid in spherical coordinates \{$R$, $\Theta$\} (150$\times$50 cells), 
assuming both azimuthal and mirror (vertically on either side of the midplane) 
symmetry.  Here, $\Theta$ is the latitude measured from the pole ($\Theta = 0$) 
to the equator (the disk midplane, $\Theta = \pi/2$).  The grid is linear in 
the $\Theta$ dimension, with a fine resolution scale at the midplane and well 
into the disk atmosphere ($\Theta \le 0.5$) and coarser sampling near the 
pole.  The grid in the radial dimension is logarithmic, running from an inner 
edge ($R_{{\rm in}}$; see \S 3.2) to an outer boundary chosen to be large 
enough to comfortably accomodate all of the disks in this sample (fixed here at 
1000\,AU, corresponding to 8\arcsec\ projected on the sky).  

The two-dimensional density structure in this coordinate system is 
\begin{equation}
\rho(R,\Theta) = \frac{\Sigma}{\sqrt{2 \pi} R h} \exp \left[- \frac{1}{2} \left( \frac{\pi/2 - \Theta}{h} \right) ^2 \right],
\end{equation}
where $\Sigma$ is the radial surface density profile, and $h$ is an angular 
scale height (in latitude).  The latter sets the width of the Gaussian vertical 
density profile, and is taken to be a power-law 
\begin{equation}
h = h_c \left( \frac{R}{R_c} \right) ^{\psi},
\end{equation}
where $h_c$ is normalized at a characteristic radius $R_c$ and $\psi$ sets the 
flaring angle of the disk.  This angular parameterization is related to a more 
intuitive physical scale height $H_R$, such that
\begin{equation}
H_R \approx R h = R h_c \left( \frac{R}{R_c} \right)^{\psi} = H_c \left( \frac{R}{R_c} \right)^{1+\psi}.
\end{equation}
We use a generic surface density profile characterized by a power-law in the 
inner disk and an exponential taper at large radii 
\citep[see][]{lyndenbell74,hartmann98},
\begin{equation}
\Sigma = \Sigma_c \left( \frac{R}{R_c} \right)^{-\gamma} \exp \left[ - \left( \frac{R}{R_c} \right)^{2-\gamma} \right].
\end{equation}
The normalization $\Sigma_c$ can be written in terms of the total disk mass 
$M_d$ by integrating equation (4) over the disk area; when $\gamma \ne 2$, 
\begin{equation}
\Sigma_c = (2-\gamma) \frac{M_d}{2 \pi R_c^2}.
\end{equation}  
This prescription for the two-dimensional density structure is fully described 
by 5 key parameters: the total disk mass ($M_d$), a surface density gradient 
($\gamma$), a characteristic radius ($R_c$), a scale height normalization 
($h_c$), and a scale height gradient ($\psi$).  

Note that by maintaining \{$h_c$, $\psi$\} as free parameters, we do not force 
the dust distribution to preserve hydrostatic pressure equilibrium in the 
vertical dimension.  For many disks, there is compelling evidence that dust 
grains have settled toward the disk midplane 
\citep{chiang01,dullemond04b,dalessio06}.  Therefore, the dust emission that we 
use to constrain structure parameters will not necessarily have the same 
vertical distribution as the gas (which is expected to be in hydrostatic 
equilibrium).  Some implications of that decision are discussed in \S 4.1.7.  
In previous efforts to model disk structure, the surface density profile is 
assumed to be a power-law, $\Sigma \propto R^{-p}$, with a sharp cut-off at an 
outer boundary, $R_{\rm out}$.  However, millimeter observations with improved 
sensitivity and resolution have shown that this power-law+cut-off behavior 
yields inconsistent results for the location of $R_{\rm out}$: the optically 
thin continuum emission indicates a significantly smaller $R_{\rm out}$ than 
the optically thick molecular line emission \citep{pietu05,isella07}.  
\citet{hughes08} demonstrated that this discrepancy may be an artifact of the 
sharp $\Sigma$ cut-off, and advocated a more gradual density taper at large 
disk radii.  A similar suggestion was made based on optical observations of 
silhouette disks in Orion \citep{mccaughrean96}.  The $\Sigma$ profile in 
equation (4) accomodates these observational findings and is grounded in a 
mathematical formalism for the viscous evolution of accretion disks (see \S 
5.1).

As Figure \ref{images} illustrates, the disks around SR 21 and DoAr 44 exhibit 
diminished continuum emission intensities out to sizable distances from the 
stellar position.  We will demonstrate in \S 4 that the same holds for the WSB 
60 disk on slightly smaller radial scales.  In these cases, we have modified 
the surface density profile in equation (4) in an effort to approximately 
reproduce the observed emission morphologies and SEDs.  We define a radius, 
$R_{\rm cav}$, such that the modified surface density $\Sigma^{\prime} = 
\delta_{\rm cav} \Sigma$ when $R \le R_{\rm cav}$, where $\Sigma$ is taken from 
equation (4) and $\delta_{\rm cav} < 1$ artificially reduces the densities 
inside the cavity (deficit) in the disk emission.

\subsection{Fixed Inputs}

Various other components of the modeling process are fixed in an effort to keep 
the problem tractable.  Of those fixed inputs, the most important are the 
properties of the dust grain population and the central star.  For the former, 
we assume a single population of spherical dust grains with a power-law 
distribution of sizes ($a$), where $n(a) \propto a^{-3.5}$ from $a_{{\rm min}} 
= 0.005$\,$\mu$m to $a_{{\rm max}} = 1$\,mm.  \citet{dalessio01} demonstrated 
that such a large maximum grain size is necessary to account for the millimeter 
continuum spectra that are typically observed for T Tauri disks \citep[see also][]{draine06}.  Opacities for this grain population were computed with a Mie 
scattering code, assuming the silicate and graphite abundances determined for 
dust grains in the interstellar medium by \citet{draine84} and their updated 
optical properties as calculated by \citet{weingartner01}.  At millimeter 
wavelengths, these opacities ($\kappa$) are similar to the widely-used 
approximation originally advocated by \citet{beckwith90}, with $\kappa \approx 
3.5$\,cm$^2$ g$^{-1}$ at 870\,$\mu$m and a spectral index $\beta \approx 1$ 
(where $\kappa_{\nu} \propto \nu^{\beta}$ in the relevant frequency range).  
The opacity spectrum used here is shown in Figure \ref{opacities}, along with 
the \citet{beckwith90} approximation and the \citet{weingartner01} interstellar 
medium opacities for reference.  

As stellar irradiation is the only heating source considered here, the 
effective temperature ($T_{\ast}$) and luminosity ($L_{\ast}$) of the central 
star are fundamental fixed inputs in the models.  To determine \{$T_{\ast}$, 
$L_{\ast}$\} in a uniform way for each individual target, we first compiled 
spectral types and foreground extinction estimates from the literature.  Using 
the \citet{kenyon95} tabulation, those spectral classifications were converted 
to $T_{\ast}$ values.  Estimates of $L_{\ast}$ were determined by scaling an 
appropriate stellar model spectrum \citep{kurucz93,hauschildt99} to match 
broadband optical photometry (see \S 4.2 for references) that was de-reddened 
according to the \citet{mathis90} extinction law.  For large foreground 
extinctions ($A_V \ge 3.5$), we used a larger value of the total-to-selective 
extinction ratio ($R_V = 5.0$, rather than the standard 3.1) that is found to 
be more appropriate in dense molecular clouds \citep[e.g.,][]{martin90}.  The 
inner radius of the disk structure grid, $R_{\rm in}$, is also set by the 
stellar properties, based on a crude approximation of where temperatures are 
high enough to sublimate the dust grains.  We define
\begin{equation}
R_{\rm in} = R_{\ast} \left(\frac{T_{\ast}}{T_s}\right)^2 = \left(\frac{L_{\ast}}{4 \pi \sigma T_s^4}\right)^{1/2},
\end{equation}
where $R_{\ast}$ is the stellar radius, $T_s$ is the sublimation temperature 
and is fixed at 1500\,K, and we assume that self-irradiation by the inner disk 
rim does not substantially contribute to the heating at $R_{\rm in}$ 
\citep{dullemond01}.  Table \ref{stars_table} lists the relevant stellar 
parameters and literature references adopted here.  Although not explicitly 
used in the modeling, we include stellar masses and ages in Table 
\ref{stars_table} that were determined with reference to the \citet{siess00} 
pre-main-sequence stellar evolution models in a Hertzsprung-Russell diagram.  

The viewing geometry of the disk, specified by the inclination and position 
angle on the sky, are required to compute synthetic data from the models.  In 
most cases, we fix the \{$i$, PA\} values based on elliptical Gaussian fits of 
the continuum visibilities (see \S 2).  Given the sensitivity and angular 
resolution of our SMA data, these estimates are generally sufficient.  For the 
few cases with clear emission line detections, we can use the resolved CO 
spatio-kinematics to test and refine those viewing geometry estimates.  We 
first make an estimate of the disk structure using the viewing geometry 
inferred from the Gaussian fits to the continuum data.  That structure model is 
then used as an input to the non-local thermodynamic equilibrium Monte Carlo 
radiative transfer code {\tt RATRAN} \citep{hogerheijde00} to determine CO 
level populations and compute synthetic visibilities at the same spatial 
frequencies and velocity channels sampled by the SMA.  We assume a Keplerian 
velocity field based on the stellar masses in Table \ref{stars_table}, a 
turbulent velocity width of 0.1\,km s$^{-1}$, and a homogeneous CO abundance of 
$\sim$10$^{-4}$-10$^{-5}$ relative to H$_2$ (varied to match the line 
intensities) in regions of the disk where $T \ge 20$\,K (to account for 
depletion onto dust grains).  The synthetic visibilities were compared with the 
SMA data over a grid of \{$i$, PA\} values to determine a refined viewing 
geometry estimate (the model fits were then re-performed with the refined 
\{$i$, PA\} values).  In the 3 relevant cases, the initial estimates of \{$i$, 
PA\} from the Gaussian fits to the continuum data were found to be accurate, 
lying within 5\degr\ for the inclination and 8\degr\ for the position angle.  
As an example, Figure \ref{AS209_CO} shows the CO channel maps for the AS 209 
disk along with the corresponding best-fit model and residuals determined in 
this way.

\subsection{Radiative Transfer}

For a given set of stellar properties, dust grain population, and density 
structure as described above, we compute the temperature structure of a model 
disk and generate synthetic data products with the two-dimensional axisymmetric 
Monte Carlo radiative transfer code {\tt RADMC} (v3.1), created and developed 
by C.~P.~Dullemond.  {\tt RADMC} utilizes an algorithm similar to the one 
described by \citet{bjorkman01} to simulate the propagation of photons through 
a dust medium that is heated solely by irradiation (in this case from the 
central star).  The basic principles of the {\tt RADMC} code were described by 
\citet{dullemond04a}.  In this version, a diffusion algorithm is employed to 
calculate temperatures in the densest regions of the disk (near the midplane of 
the inner disk), where photon statistics are poor in the Monte Carlo 
simulations.  With the results of the radiative transfer calculations, a 
raytracing program is used to generate a model SED and millimeter continuum 
visibilities (at the same spatial frequencies \{$u$, $v$\} that were sampled by 
the SMA observations) for a given viewing geometry.  To simulate the smearing 
effect of atmospheric phase noise, the model visibilities are convolved with a 
0\farcs1 Gaussian ``seeing" kernel.  The small seeing disks inferred here have 
a minimal impact on the visibilities, but are included for completeness.

\subsection{Estimating Model Parameters}

Our modeling philosophy is based on finding the disk structure parameters that 
can best reproduce both the de-reddened broadband SED and the SMA continuum 
visibilities.  In practice, this is achieved by varying the 5 free parameters 
that describe the density structure, \{$M_d$, $\gamma$, $R_c$, $h_c$, $\psi$\}, 
such that the total $\chi^2 = \chi^2_{\rm sed} + \chi^2_{\rm vis}$ statistic is 
minimized (two additional parameters, \{$R_{\rm cav}, \delta_{\rm cav}$\}, are 
varied for the SR 21, WSB 60, and DoAr 44 disks).  The $\chi^2_{\rm vis}$ 
values are computed as the sum of the $\chi^2$ values for the real and 
imaginary components of the visibilities at each individual spatial frequency, 
with the errors based on the visibility weights and amplitude calibration 
uncertainty \citep[e.g.,][]{lay97,guilloteau98}.  The broadband SEDs are 
constructed from the literature (see \S 4.2 for references in individual 
cases), and uncertainties at each frequency are computed as the quadrature sum 
of standard errors and absolute calibration uncertainties.  Data shortward of 
0.5\,$\mu$m ($V$-band) and near 10\,$\mu$m are excluded so that the fits are 
not affected by excess ultraviolet emission from accretion and the detailed 
shape of silicate emission profiles, respectively.  The parameter estimation 
process begins by computing $\chi^2$ values over a broad, coarse grid of 
parameter-space.  Those results guide a refined search for a global minimum 
$\chi^2$ using the downhill simplex algorithm {\tt AMOEBA} \citep{press92}.  
For the SR 21, WSB 60, and DoAr 44 disks, we explore an abridged 
parameter-space, with a focus on estimating a size for the central emission 
cavity ($R_{\rm cav}$): more detailed modeling will be treated elsewhere 
\citep[e.g., see][]{brown09}.

Because of the finite number of simulated photon packages used in Monte Carlo 
radiative transfer calculations, the output synthetic data products contain 
some inherent noise.  In general, this noise can be effectively suppressed with 
a sufficiently large number of photon packages and the line-of-sight averaging 
employed in the post-processing raytracing program.  However, there is some 
concern about the Monte Carlo noise manifested in the temperatures near the 
disk midplane $-$ where the photon statistics are relatively poor in the 
simulations $-$ because of their direct impact on the synthetic millimeter 
visibilities (recall $S_{\nu} \propto \kappa \Sigma T$).  To assess the 
influence of that noise on the local shape of $\chi^2$-space used in the {\tt 
AMOEBA} minimization algorithm, we performed two ``repeatability" tests.  
First, we conducted {\tt RADMC} + raytracing calculations multiple times for 
fixed parameter sets and compared the resulting synthetic data products with 
the real data and errors.  These comparisons indicated that the Monte Carlo 
noise contributes only minimally to the $\chi^2$ values, significantly less 
than the observational uncertainties.  And second, we initialized the {\tt 
AMOEBA} algorithm with a variety of different parameter sets and compared the 
end results of the minimization process (i.e., the best-fit parameter values).  
Those results were not significantly different from one another, again 
indicating that the Monte Carlo noise does not have an adverse affect on the 
local shape of $\chi^2$-space. 

While a simple $\chi^2$ sum for the independent datasets (visibilities and SED) 
is the appropriate goodness-of-fit statistic from a purely mathematical 
viewpoint, it is important to highlight the relative impact of each dataset on 
the fitting process.  For a given disk, there are many more visibilities than 
SED datapoints ($N_{\rm vis}/N_{\rm sed} > 10^3$).  However, the 
signal-to-noise ratios for individual visibilities are substantially lower than 
for the SED fluxes, such that the relative contribution of each visibility 
point to the total $\chi^2$ value is roughly a factor of $\sim$$10^3$ less than 
that for each individual SED point (within an order of magnitude).  In 
practice, this balance between the quantity and quality of individual 
datapoints ensures that the total $\chi^2$ value is appropriately tempered if 
$\chi^2_{\rm sed}$ becomes too large, despite the fact that the large 
$\chi^2_{\rm vis}$ values generally have more influence over the minimization 
algorithm.  This balance is just a fortuitous coincidence, but future 
interferometric datasets will offer improved sensitivity and vastly increased 
$N_{\rm vis}$.  To continue with this kind of multi-dataset modeling effort 
with superior millimeter data (e.g., from the {\it Atacama Large Millimeter 
Array}), the relative weighting of datasets will need to be re-evaluated.  To 
that end, we experimented with two alternative ways of defining a 
goodness-of-fit statistic.

In one case, we minimized the sum $\tilde{\chi}^2_{\rm vis} + 
\tilde{\chi}^2_{\rm sed}$ of the {\it reduced} $\chi^2$ values, defined as 
$\tilde{\chi}^2 = \chi^2/\nu$ where $\nu = N - 5$ is the degrees of freedom in 
the fit (the number of datapoints less the number of free parameters in the 
model).  However, the relative weightings (i.e., signal-to-noise ratio, SNR) 
for each dataset imply that $\tilde{\chi}^2_{\rm sed} > 
\tilde{\chi}^2_{\rm vis}$ in general, driving the minimization to parameter 
sets that provide poor matches with the millimeter data.  The large 
$\tilde{\chi}^2_{\rm sed}$ values are usually due to relatively small 
mismatches with the high SNR optical/infrared fluxes, where the detailed 
stellar properties and heating/scattering near the inner disk rim are important 
(and not treated in detail here).  Therefore, parameter estimation based on 
this $\tilde{\chi}^2$ statistic is too much influenced by physical conditions 
near the star-disk interface that are not of interest here.  One could imagine 
an additional re-weighting scheme as a function of SED wavelength, but it is 
not clear how to assign those weights nor how such manipulations could affect 
the best-fit parameter estimates.  In a separate experiment, we instead aimed 
to balance the impact of the two datasets using azimuthally-averaged visibility 
profiles (see \S 4.1 for details) such that $N_{\rm vis} \sim N_{\rm sed}$.  
However, the SNR issues persist in this definition and again lead to fits 
dominated by highly weighted short-wavelength SED points.  In the end, we adopt 
the simple summed $\chi^2$ statistic initially defined above because it imposes 
the fewest assumptions upon the data (and still produces quality fits).  
Similar modeling in the future will likely require some effort to distill the 
most useful set of information from larger and more sensitive visibility 
datasets, perhaps akin to some of the clever geometric techniques developed to 
characterize scattered light images \citep[e.g.,][]{glauser08,pinte08}.

Although this grid search and {\tt AMOEBA} minimization technique is able to 
find a minimum $\chi^2$ value in the 5-dimensional disk structure 
parameter-space, it does not provide sufficient information to accurately gauge 
uncertainties on the best-fit parameter estimates.  The computation time 
required to perform the Monte Carlo radiative transfer calculations outlined 
above is the limiting factor that makes proper estimates of uncertainties 
(e.g., from a sufficiently sampled $\chi^2$ grid search) prohibitive for such a 
large sample.  Since the shape of the surface density profile is of 
considerable interest, we gauged how accurately we can determine $\gamma$ by 
repeating the fitting process described above for various fixed $\gamma$ values 
around the best-fit estimate.  In general, we find that the fit quality is 
significantly diminished when these explored $\gamma$ values deviate from the 
best-fit value by $\sim$0.2-0.3.  A representative example of this process is 
discussed in \S 4.1.2.  While these are only qualitative estimates of the 
uncertainties on $\gamma$, they are in reasonable agreement with more 
statistically robust error estimates for similar studies 
\citep[e.g.,][]{pietu07,isella09}.

\section{Results}

\subsection{Disk Structures}

The density structure parameter values that best reproduce the data for the six 
disks with continuous emission distributions are presented in Table 
\ref{structure_table}; those for the remaining three disks with central 
emission cavities are shown separately in Table \ref{trans_table}.  Also listed 
are the fixed values for the inclinations, major axis position angles, and 
inner disk radii (Table \ref{structure_table} only; see \S 3.2), as well as the 
$\tilde{\chi}^2$ statistics corresponding to the visibility and SED datasets 
separately (see \S 3.4).  The scale height parameter $h_c$ has been recast into 
a more common spatial (rather than angular) format, as described in equation 
(3).  The distributions of the model parameters are shown together in Figure 
\ref{histograms}.  In the following sections, we discuss each of these 
parameters in the context of the sample as a whole.  Commentaries on the 
results for individual disks are provided in \S 4.2.  The synthetic data 
produced by the best-fit structure models are compared with the observations in 
Figures \ref{results} and \ref{trans}.  From left to right in these figures, 
each row shows the observed millimeter continuum image (as in Figure 
\ref{images}), the best-fit model image synthesized in the same way as the 
data, the imaged residuals, the broadband SED, and the elliptically-averaged 
variation of the real part of the visibilities as a function of the deprojected 
interferometer baseline length.  

The ``visibility profiles" in the rightmost panels of Figures \ref{results} and 
\ref{trans} conveniently display the continuum emission at all of the spatial 
scales that are sampled by the interferometer.  The profiles were generated by 
averaging the visibilities in 20\,k$\lambda$ annular bins in a coordinate 
system that accounts for the disk viewing geometry.  The abscissae mark the 
deprojected baseline length in this coordinate system, ${\cal R}_{uv} = 
(d_a^2 + d_b^2)^{1/2}$, where the major and minor axes are $d_a = (u^2 + 
v^2)^{1/2} \sin{\phi}$ and $d_b = (u^2 + v^2)^{1/2} \cos{\phi} \cos{i}$, 
respectively \citep{lay97}.  Here \{$u$, $v$\} are the Fourier spatial 
frequency coordinates and $\phi = \arctan{(v/u)} - {\rm PA}$.  The ordinates 
show the real part of the visibility flux (averaged in each bin), calculated 
with an appropriate phase shift to compensate for the centroid position offset 
relative to the observed phase center.  For a circularly symmetric disk that is 
both vertically and optically thin, the visibility profile constructed in this 
way represents the Fourier transform of the radial surface brightness 
distribution.  In the Figure \ref{results} and \ref{trans} images, crosshairs 
mark the centroid position and major axis position angle; their relative 
lengths denote the aspect ratio set by the disk inclination.  The model fits 
are overlaid in red for the SEDs and visibility profiles, and the contributions 
of the stellar photospheres are shown as blue dashed curves.

Note that we have included higher resolution inset images for the WSB 60 disk 
in Figure \ref{trans}.  Those images were generated from the same visibility 
data, but with increased weight on the longer baselines (with robust = $-$1 and 
excluding baselines $<$40\,k$\lambda$) to accentuate the continuum structure at 
small disk radii.  The synthesized beam in these insets has dimensions of 
$0\farcs42\times0\farcs22$ at PA = 21\degr\ ($\sim$2$\times$ higher resolution 
in the short axis compared to the main image).  Despite the symmetric 
appearance of the WSB 60 disk in the main panel at lower resolution, the inset 
clearly demonstrates that the emission is not centrally peaked.  This behavior 
was the motivation for using the disk cavity density model in this case.  
Although some low-level asymmetries are apparent in the data images (e.g., for 
GSS 39 or AS 209), no significant ($\ge 3\sigma$) deviations from the assumed 
symmetric models are found in the residual images.  As with the model fits, the 
data$-$model subtraction is performed on the visibilities, and not in the image 
plane.  Those residual visibilities are then Fourier inverted, deconvolved, and 
restored with the synthesized beam in the same way as for the data and model 
(see \S 2).  The absence of significant residual emission associated with these 
apparent asymmetries implies that their origin is consistent with noise.  
Comparisons of images made with alternative weightings of the visibility data 
in the deconvolution process support that conclusion.  

In the following sections, we discuss each of the key free parameters in these 
structure models, their values for the sample as a whole, and their distinct 
signatures on the observational data.  There are three important points to keep 
in mind: (1) the amount of millimeter continuum emission scales with \{$M_d$, 
$h_c$\}; (2) the spatial distribution of that emission is determined by 
\{$\gamma$, $R_c$, $\psi$\}; and (3) in practice, the shape of the infrared SED 
plays a key role in determining the vertical structure parameters \{$h_c$, 
$\psi$\}, leaving \{$M_d$, $\gamma$, $R_c$\} to be constrained by the 
millimeter data.

\subsubsection{Disk Masses ($M_d$)}

As shown in Figure \ref{histograms}a, the total masses inferred for the sample 
disks range from 0.005-0.14\,M$_{\odot}$ (assuming a 100:1 gas-to-dust mass 
ratio).  The disks around SR 21, WSB 60, and DoAr 44 are associated with the 
lowest $M_d$ values (0.005-0.02\,M$_{\odot}$), due to the diminished densities 
in the central regions required to explain the observed emission morphologies.  
But regardless, the masses for this sample lie at the high end of the combined 
$M_d$ distribution for Taurus and Ophiuchus disks \citep[ranging from 
$\sim$0.0001-0.1\,M$_{\odot}$; see Fig.~10 in][]{aw07b}, highlighting an 
important sample selection bias.  To obtain sufficient sensitivity on long SMA 
baselines, the sample disks were chosen to be among the brightest 850\,$\mu$m 
sources in the Ophiuchus star-forming region.  Because of the low optical 
depths at these wavelengths, bright emission is associated with higher disk 
masses \citep{beckwith90}.  Although this sample may not be representative of 
the median $\sim$1\,Myr-old disk, the $M_d$ values inferred here are comparable 
to or substantially larger than the minimum mass of the primordial solar nebula 
\citep[$\sim$0.01\,M$_{\odot}$;][]{hayashi81}.  Therefore, the sample includes
good examples of the disks that are most capable of making planets, at least in 
terms of containing a sufficient amount of raw material.  However, their true 
potential for planet formation depends intimately on how that material is 
distributed spatially.

\subsubsection{Surface Density Gradients ($\gamma$)}

The distribution of mass in the disk is characterized by the parameter 
$\gamma$, which sets the shape of the surface density profile.  Figure 
\ref{histograms}b demonstrates that the disks in this sample have similar 
$\Sigma$ profiles, with a narrow range of gradients ($\gamma \approx 0.4$-1.0) 
around the median value, $\gamma = 0.9$.  That typical value corresponds to a 
surface density profile that varies roughly inversely with radius in the inner 
disk ($\Sigma \propto 1/R$ when $R \lesssim R_c$) and smoothly merges into a 
steeper exponential decrease at larger radii ($\Sigma \propto 1/e^R$ when $R 
\gtrsim R_c$).  Figure \ref{sigma} exhibits the inferred $\Sigma$ profiles, 
with the dust surface densities utilized in the modeling scaled up 100$\times$ 
to represent the total gas+dust densities.  As a reference, the dark gray 
regions plotted in Figure \ref{sigma} show the MMSN surface densities for 
Saturn, Uranus, and Neptune determined by \citet{weidenschilling77}.  The 
radial widths of those regions mark the annular extent over which the augmented 
planetary masses were spread, and their heights denote the uncertainties in the 
chemical composition of the planets relative to cosmic abundances.  A more 
detailed comparison is made in \S 5.2.

Referring back to the qualitative emission-structure relation, $S_{\nu} \propto 
\kappa \Sigma T$, we note that $\gamma$ should have a direct impact on the 
morphology of the millimeter continuum emission.  An examination of Figure 
\ref{sigma} demonstrates that the inner disk $-$ where $\Sigma \propto 
1/R^{\gamma}$ $-$ is generally not well-resolved, even with the maximal 
resolution scale afforded by the SMA data (marked in light gray).  Therefore, 
the observed emission morphologies are most sensitive to the shape (and 
location) of the $\Sigma$ taper in the outer disk, where $\Sigma \propto 
1/\exp{(R^{2-\gamma})}$.  Such behavior implies that smaller $\gamma$ values 
produce more centrally-concentrated emission morphologies, or steeper surface 
brightness profiles.  It is worthwhile to point out that this is exactly the 
opposite relationship between the emission distribution and the gradient $p$ in 
the standard power-law assumption for surface densities, where $\Sigma \propto 
R^{-p}$ out to some cut-off radius \citep[e.g.,][]{aw07}.  

This relation between $\gamma$ and the emission morphology is manifest in 
Figure \ref{gamma_unc}, which serves as a qualitative demonstration of how 
accurately we can measure $\gamma$ values (see \S 3.4 for details).  In this 
figure, we show the SED and visibility profile data for the WaOph 6 disk (as in 
Figure \ref{results}) with the best-fit $\gamma = 1.0$ models overlaid in red.  
The blue and green curves show the models that best reproduce the data for 
$\gamma$ values fixed at 0.8 and 1.2, respectively.  All three of these models 
match the SED well, but there are clear differences between them in a 
comparison with the observed visibilities.  Because the data probe a steeper 
$\Sigma$ taper for the $\gamma = 0.8$ (blue) model compared to $\gamma = 1.0$ 
(red), we observe a steeper brightness profile with more emission concentrated 
on smaller spatial scales.  In terms of the visibility profiles, this results 
in the $\gamma = 0.8$ model over-predicting the amount of emission on the 
longest baselines.  The exact opposite is true for the $\gamma = 1.2$ (green) 
model.  We should emphasize that while Figure \ref{gamma_unc} demonstrates the 
connection between $\gamma$ and the observations, it is only representative of 
a qualitative exploration of parameter space, and not a formal statistical 
characterization of the uncertainties on $\gamma$.

\subsubsection{Characteristic Radii ($R_c$)}

Figure \ref{histograms}c shows that the characteristic radii that govern where 
the $\Sigma$ profiles transition from power-laws to exponential tapers lie in 
the range $\sim$20-200\,AU for the sample disks.  A larger $R_c$ generally 
means more emission on larger spatial scales, although the role of this 
parameter as a fulcrum in the $\Sigma$ and $h$ radial profiles can complicate 
that behavior.  Since these characteristic radii will play a key role in a 
discussion of viscous evolution models (\S 5.1), it is worth reiterating that 
they are not the same as the standard cut-off radius ($R_c \ne R_{\rm out}$).  
As mentioned in \S 3.1, sharp outer edges are disfavored by resolved optical 
observations \citep{mccaughrean96} and a comparison between millimeter 
continuum and CO spectral images \citep{hughes08}.  Moreover, the new 
continuum data presented here show no evidence for the visibility nulls that 
would be produced by such sharp edges (with the notable exceptions of the disks 
around SR 21, WSB 60, and DoAr 44).

\subsubsection{Vertical Structure ($h_c$ and $\psi$)}

Recall that the millimeter continuum emission from a disk scales with the 
product of density and temperature ($S_{\nu} \propto \kappa \Sigma T$).  In our 
models, the temperature structure is determined by how effectively the disk 
material intercepts and reprocesses energy from the star.  Most of that energy 
is absorbed by dust grains at some height above the disk midplane, creating an 
intimate connection between the vertical structure of the disk and its 
temperature distribution.  In a flared disk, a large surface area of dust can 
be directly irradiated by the star.  Therefore, more energy will be absorbed in 
the disk atmosphere and subsequently re-emitted, both into space and deeper 
into the disk interior.  In essence, vertically extended (larger $h_c$) disks 
are heated more efficiently, emit more from their atmospheres, and have higher 
interior temperatures.  Consequently, a larger $h_c$ produces brighter emission 
at both infrared and millimeter wavelengths.  By controlling the vertical 
extent of the disk at different radii, the scale-height gradient $\psi$ 
determines how stellar energy is deposited into the disk atmosphere as a 
function of radius.  Steeply flaring disks (larger $\psi$) will re-emit more of 
that energy from larger radii, both outward toward an observer and inward to 
heat the disk interior.  Since cooler dust preferentially emits at longer 
wavelengths, a larger $\psi$ will generate a flatter infrared SED.  Likewise, a 
larger $\psi$ results in a flatter temperature profile in the disk interior, 
and therefore a more diffuse millimeter emission morphology (a flatter surface 
brightness profile). 

The histograms in Figure \ref{histograms}d,e show scale heights from 4-20\,AU
(at a fiducial radius of 100\,AU) and $\psi$ values from 0.04-0.26 for the
sample disks.  While these numerical ranges are small, they represent a diverse
set of vertical structures $-$ from flat and cold (e.g., WaOph 6) to flared and 
warm (e.g., AS 205).  Figure \ref{Tmid} shows the radial temperature profiles 
at the disk midplanes, derived from the radiative transfer calculations (\S 
3.3; for clarity, the disks with central cavities are not shown).  These 
profiles generally behave as power-laws, $T \propto R^{-q}$ with $q \approx 
0.5$-0.6 over a wide range of radii, from just beyond $R_{\rm in}$ out to 
$\sim$$R_c$.  At larger radii, the exponential $\Sigma$ taper facilitates a 
more efficient heating of the disk interior, resulting in a flatter midplane 
$T$ profile ($q \approx 0.3$-0.4).  The disks with surface density 
discontinuities (around SR 21, WSB 60, and DoAr 44) have $T$ profiles with a 
distinct kink and hotter temperatures around $R_{\rm cav}$, marking the 
rarefied dust regions just interior to where the higher density outer disk is 
frontally illuminated by the star.

\subsubsection{Central Cavity Properties ($R_{\rm cav}$ and $\delta_{\rm cav}$)}

To reproduce the diminished millimeter emission in the central regions of the 
disks around SR 21, WSB 60, and DoAr 44, we introduced two additional 
parameters \{$R_{\rm cav}$, $\delta_{\rm cav}$\} that act to decrease the 
surface densities inside a disk cavity.  The scaled-down densities of warm dust 
inside the cavity reduce the amount of infrared excess emission, producing a 
deficit in the SED compared to a disk with a continous density distribution.  
The location of the cavity edge, $R_{\rm cav}$, determines the wavelength where 
the SED deficit recovers; larger $R_{\rm cav}$ values produce SED deficits out 
to longer infrared wavelengths (corresponding to cooler temperatures).  The 
depth of the SED deficit is set by the parameter that scales down the 
densities, $\delta_{\rm cav}$; a smaller value (an emptier cavity) generates 
less emission and therefore a deeper deficit.  The diminished $\Sigma$ values 
also lead to a proportional decrease in the millimeter emission inside the 
cavity, producing a ring-like morphology (see Fig.~\ref{trans}).  The sharp 
density change at the cavity edge generates a null in the millimeter 
visibilities at a spatial frequency associated with the edge location, 
$R_{\rm cav}$ \citep{hughes07}; larger $R_{\rm cav}$ values produce nulls at 
shorter baselines.  The density contrast in the cavity sets the emission levels 
on small spatial scales, such that a lower $\delta_{\rm cav}$ produces a more 
pronounced null.  

In practice, our modeling relies heavily on the location and depth of that null 
to determine the values of \{$R_{\rm cav}$, $\delta_{\rm cav}$\}, due to the 
relatively sparse infrared SED coverage for these disks.  Those parameters are 
expected to be degenerate with the structure parameters of the remnant outer 
disk because of the short spatial dynamic range available in that region.  
Moreover, the cavity parameters derived here are only appropriate for the 
simple model assumptions that were made in \S 3.1.  Their values would be 
modified depending on the details of the cavity edge \citep[e.g., an irradiated 
``wall";][]{dalessio05,hughes09} or the dust content in the cavity interior.  
For example, others have typically adopted smaller dust grain sizes ($a_{\rm 
max} \sim 1$-10\,$\mu$m) inside disk cavities to more faithfully reproduce 
infrared spectra \citep[e.g.,][]{calvet05} and accomodate calculations that 
track inward dust filtration from the outer disk \citep{rice06}.  Compared to 
the grain size distribution adopted here, those smaller grains have opacities 
that are lower at millimeter wavelengths and higher in the infrared.  Even 
meager densities of such small dust grains inside the cavity can produce 
substantial infrared emission without changing the millimeter continuum.  
Perhaps this is a solution that would better reproduce the observations for a 
case like the DoAr 44 disk.  Alternatively, a ``gap" model, where the cavity 
contains a detached remnant of the inner disk, could generate similar 
observational signatures \citep[e.g.,][]{espaillat07,espaillat08}.

\subsubsection{Comments on Parameter Degeneracies}

In practice, the different structure parameters used here can have similar 
effects on the data, leading to some basic model degeneracies.  For example, 
$M_d$ and $h_c$ have the same scaling effect on the amplitude of millimeter 
emission, while $\gamma$ and $\psi$ have similar impacts on how that emission 
is distributed spatially.  Fortunately, those degeneracies are alleviated by 
requiring the models to fit the SED and millimeter visibilities jointly.  For 
instance, unlike the surface density parameters \{$M_d$, $\gamma$\}, the scale 
height parameters \{$h_c$, $\psi$\} can be constrained through their strong 
impact on the infrared SED.  However, some model degeneracies remain, perhaps 
most notably between $\gamma$ and $R_c$.  \citet{mundy96} first discussed an 
analogous degeneracy between emission gradients and size scales, demonstrating 
that the data can sometimes be explained equally well by trading off steeper 
gradients and larger sizes \citep[see also][]{aw07}.  In our models, this 
degeneracy is slightly complicated due to the form of the $\Sigma$ profile 
adopted in \S 3.1.  For low $\gamma$ values, the surface density (and therefore 
emission) profile is similar to a power-law + cutoff model; therefore, the data 
can accomodate larger $\gamma$ values by increasing $R_c$.  However, as 
$\gamma$ gets larger the data are increasingly insensitive to $R_c$, and its 
precise value becomes difficult to determine.  The experiments to estimate how 
well $\gamma$ can be constrained (see \S 3.4 and \S 4.1.2) indicate that this 
degeneracy is quantitatively weaker than in previous studies, due to the 
improved spatial dynamic range of the data.

\subsubsection{Comments on Opacities}

All estimates of disk structure parameters are degenerate with the assumed 
opacities.  There are two distinct structure-opacity relationships that merit 
attention.  First, and perhaps most straightforward, is a density-opacity 
degeneracy related to interpreting the millimeter continuum emission.  Because 
that emission is optically thin, it is sensitive to the optical depth, the 
product $\kappa \Sigma$.  For a given brightness, lower opacities would lead us 
to infer higher densities, and vice versa.  Likewise, spatial opacity 
variations would affect constraints on the shape of the $\Sigma$ profile 
($\gamma$).  For example, a radial opacity gradient might be a natural outcome 
of the grain growth process \citep[e.g.,][]{dullemond05,garaud07}.  Higher 
densities and velocities in the inner disk would decrease growth timescales 
compared to larger radii, leading to an opacity profile that increases with 
radius.  By assuming a spatially uniform $\kappa$, we could underestimate 
$\gamma$ in the inner disk.  Disentangling this density-opacity ambiguity is a 
formidable challenge, but some progress is feasible by exploiting the shape of 
the millimeter continuum spectrum \citep[e.g.,][]{beckwith91,rodmann06}.  In 
the future, spatially resolved millimeter SEDs could help characterize the 
relative grain growth efficiencies at different locations throughout the disk. 

A second degeneracy between the vertical structure of the disk and the opacity 
is more difficult to quantify in a generic way.  The energy from stellar 
radiation is deposited in the disk atmosphere, at a location that depends on 
the optical/infrared opacities.  If those opacities are low, stellar radiation 
can penetrate deeper into the disk before it is absorbed and then re-processed, 
and vice versa.  Therefore, a given SED can be reproduced by various 
combinations of vertical structures and material opacities.  In our models, we 
have fixed the opacities and varied some vertical structure parameters to 
reproduce the SEDs.  Some studies have taken a different approach, effectively 
fixing the vertical structure (to be in hydrostatic equilibrium) and varying 
the opacities as a function of height in the disk \citep[e.g.,][]{dalessio06}.  
In either case, the opacities are tied to the vertical structure, and therefore 
the disk temperature structure (\S 4.1.4).  Because the millimeter continuum 
emission depends on $\kappa \Sigma T$, the uncertain opacities in the disk 
atmosphere can indirectly impact our constraints on $\Sigma$.  This degeneracy 
can potentially be alleviated using independent constraints on the vertical 
structure of the disk, either with multi-transition molecular line data 
\citep[e.g.,][]{dartois03} or high resolution scattered light images 
\citep[e.g.,][]{pinte08}.

\subsection{Commentary on Individual Disks}

\begin{figure}
\epsscale{1.0}
\plotone{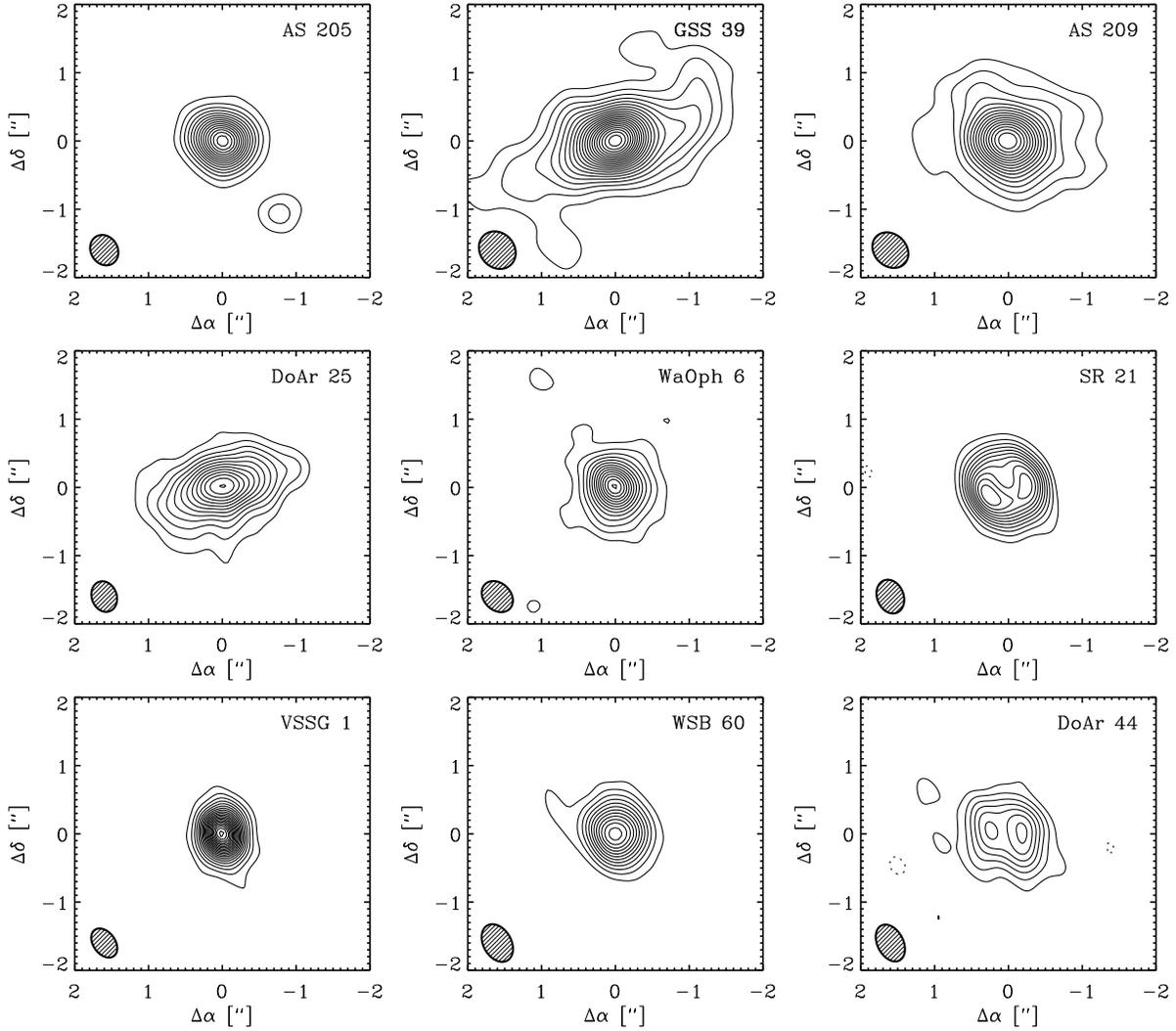}
\figcaption{Aperture synthesis images of the 870\,$\mu$m continuum emission 
from the 9 sample disks.  Each panel is 4\arcsec\ (500\,AU) on a side.  
Contours are shown at 3\,$\sigma$ intervals (rms uncertainties in Table 
\ref{data_table}).  The synthesized beams are shown in the lower left of each 
panel.  Note the detection of a disk around the AS 205 B system in the top left 
panel (see \S 4.2), as well as the prominent cleared central regions for the 
disks around SR 21 and DoAr 44.  \label{images}}
\end{figure}

\begin{figure}
\epsscale{1.0}
\plotone{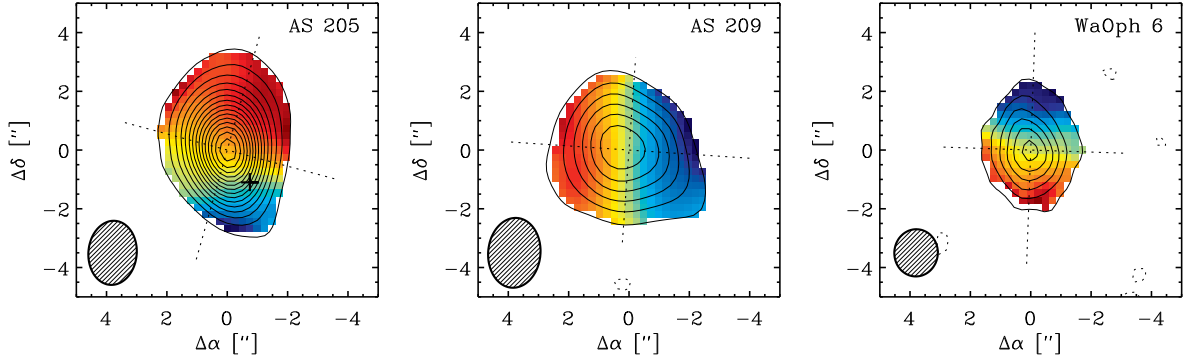}
\figcaption{Moment maps of the CO $J$=3$-$2 emission from the sample disks with 
minimal contamination from the surrounding molecular cloud.  Each panel is 
10\arcsec\ (1250\,AU) on a side.  The contours mark the velocity-integrated 
intensity (zeroth moment), with levels at 3\,$\sigma$ intervals.  The 
color-scale maps represent the intensity-weighted velocities (first moment), 
ranging from LSR velocities of 3-5\,km s$^{-1}$ for AS 205, $-3$-0\,km s$^{-1}$ 
for AS 209, and $-3$-1\,km s$^{-1}$ for WaOph 6.  The synthesized beams are 
shown in the lower left of each panel.  The location of AS 205 B is marked with 
a cross.  \label{moments}}
\end{figure}

\begin{figure}
\epsscale{0.5}
\plotone{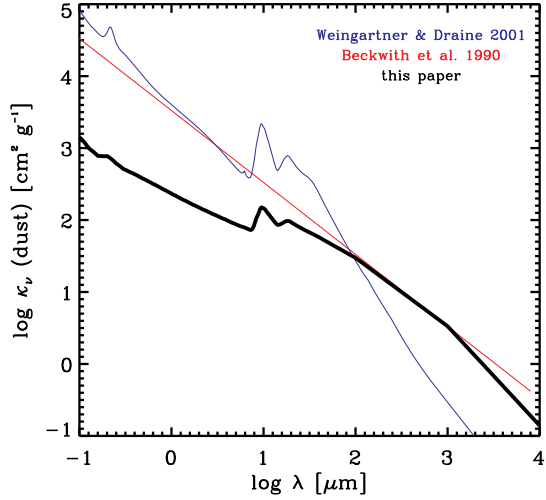}
\figcaption{The opacity spectrum for the dust grain population adopted in the 
modeling ({\it black}).  For comparison, we show the opacities for the standard 
\citet{beckwith90} prescription for disks ({\it red}) and the diffuse 
interstellar medium \citep[{\it blue};][]{weingartner01}.  \label{opacities}}
\end{figure}

\begin{figure}
\epsscale{1.0}
\plotone{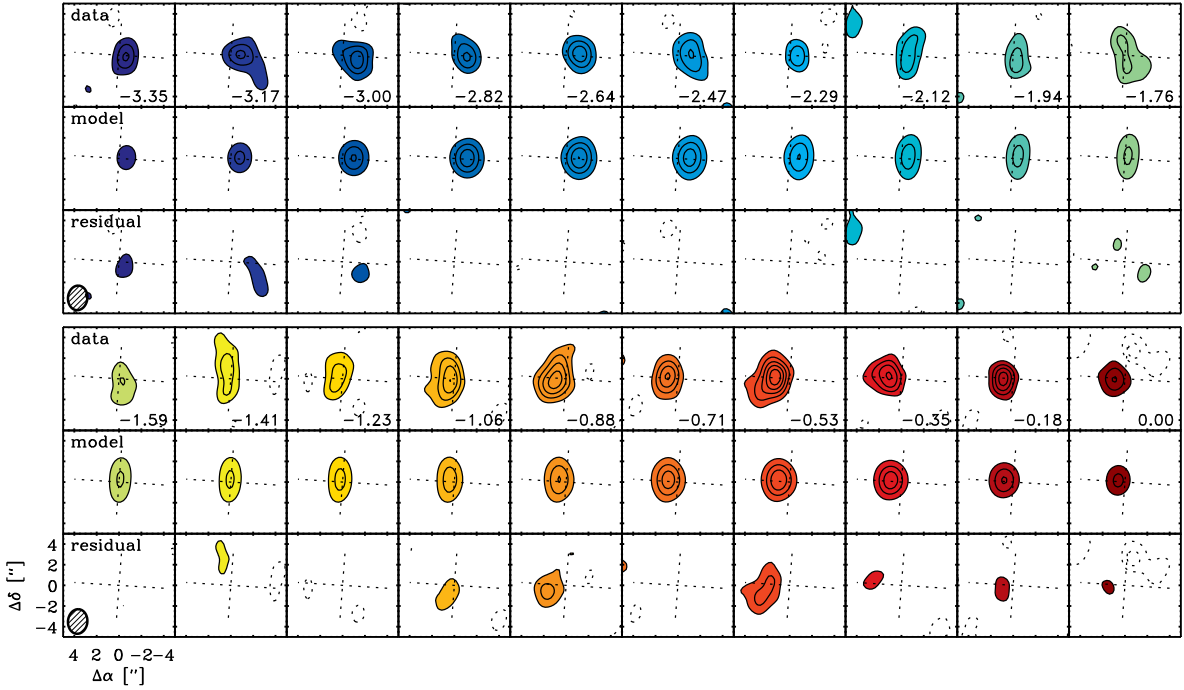}
\figcaption{CO $J$=3$-$2 channel maps for the AS 209 disk.  Each panel is 
10\arcsec\ (1250\,AU) on a side and represents a velocity width of 0.18\,km 
s$^{-1}$, as marked in the lower right corner.  Contours are shown at 
3\,$\sigma$ intervals.  The data are shown in the top rows (first and fourth), 
the best-fit model (see \S 3.2) in the middle rows (second and fifth), and the 
residuals in the bottom rows (third and sixth).  These CO data were used only 
to constrain the viewing geometry of the disk.  \label{AS209_CO}}
\end{figure}

\begin{figure}
\epsscale{1.0}
\plotone{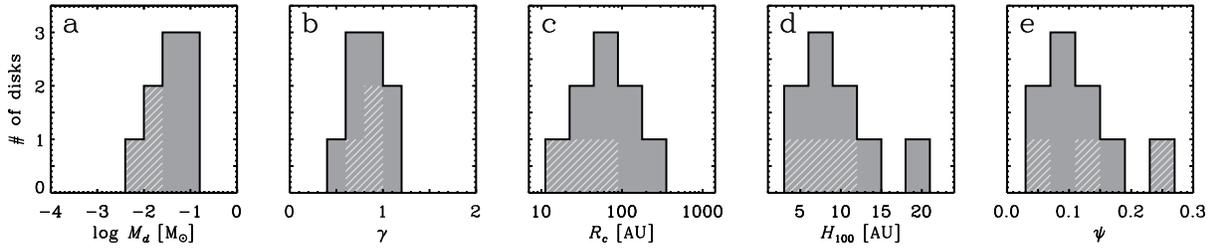}
\figcaption{Distributions of the five key structure parameters derived for the 
disks in this sample.  From left to right are the disk masses ($M_d$), radial 
surface density gradients ($\gamma$), characteristic radii ($R_c$), scale 
heights at 100\,AU ($H_{100}$; i.e., $h_c$), and the radial scale height 
gradients ($\psi$).  The contributions of the three disks with cleared central 
cavities are overlaid with hatched regions.  \label{histograms}}
\end{figure}

\begin{figure}
\epsscale{1.0}
\plotone{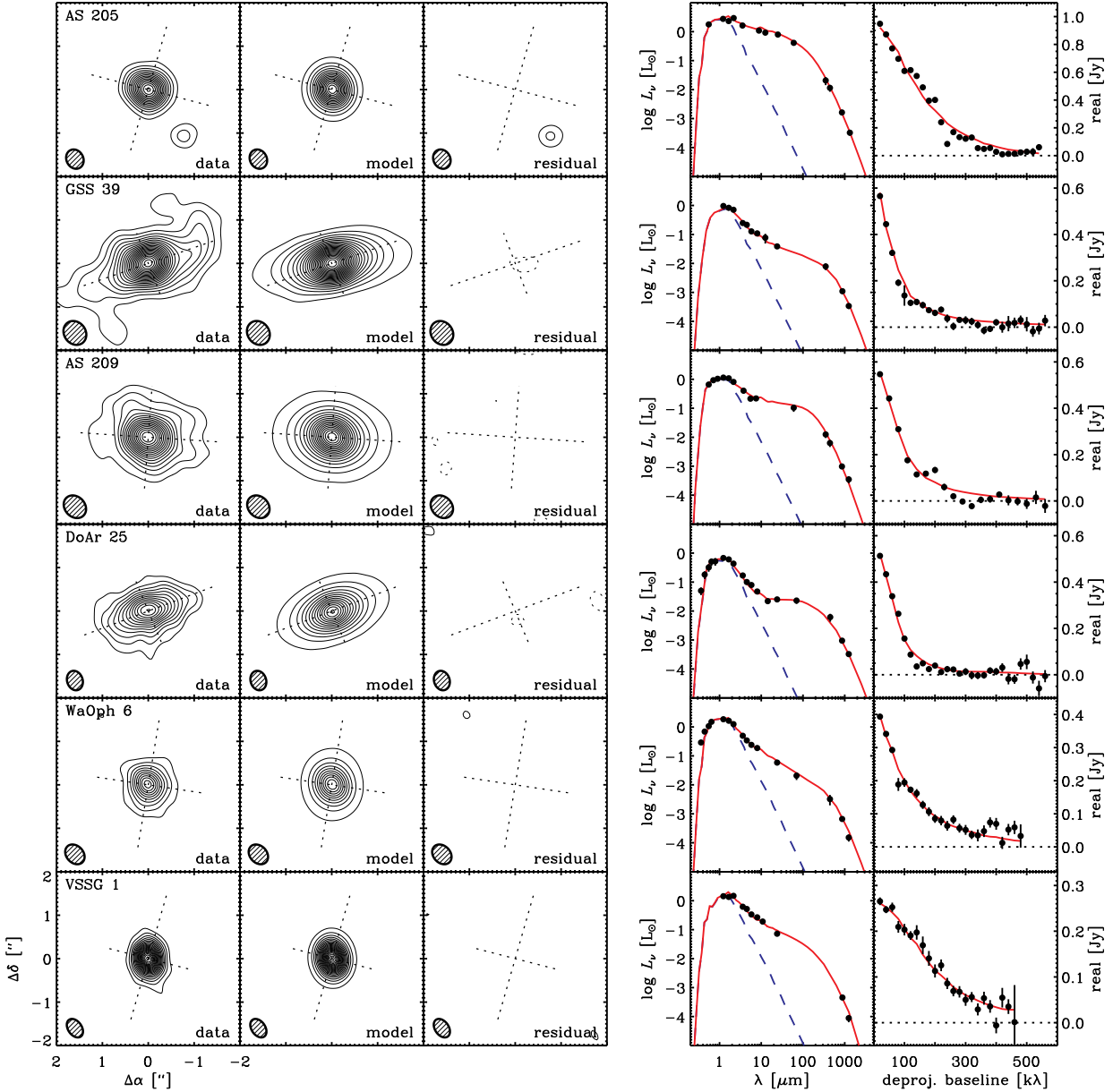}
\figcaption{Comparison of the disk structure model fits and the data.  The left 
panels show the SMA continuum image, corresponding disk model, and residuals 
(data$-$model) as described in Figure \ref{images}.  Crosshairs mark the disk 
centers and major axis position angle; their relative lengths represent the 
disk inclination.  The right panels show the broadband SEDs and deprojected 
visibility profiles (see \S 4.1 for details), with best-fit models overlaid in 
red.  The SED contributions from the stellar photosphere are shown as blue 
dashed curves.  \label{results}}
\end{figure}

\begin{figure}
\epsscale{1.0}
\plotone{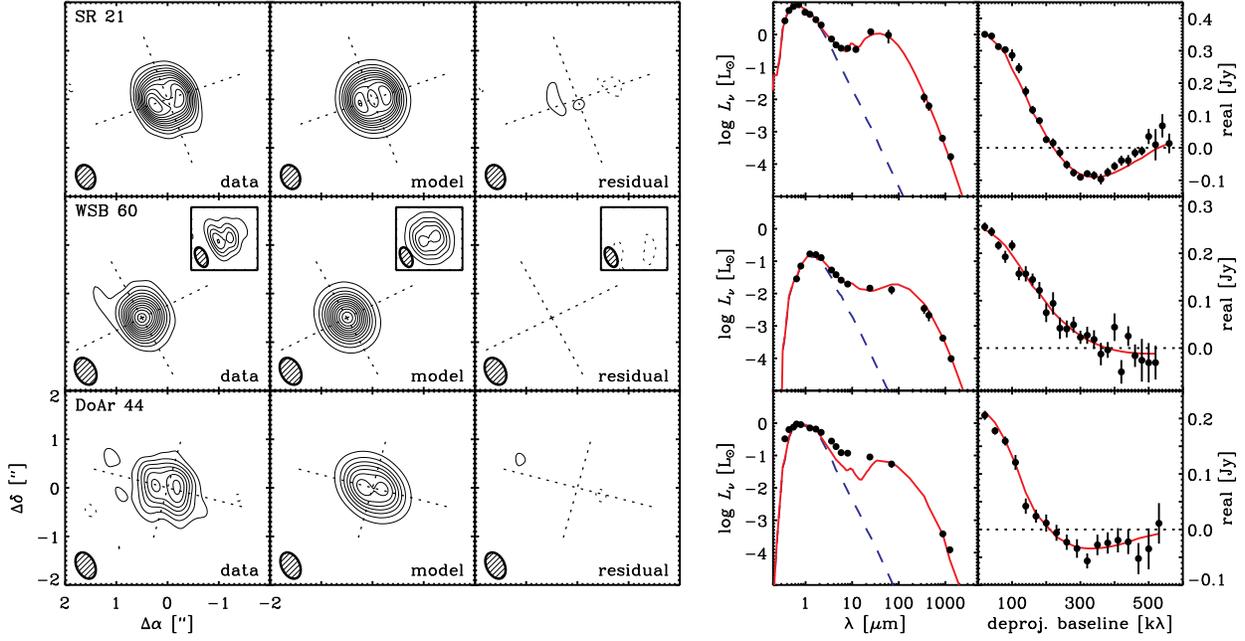}
\figcaption{Comparison of the disk structure model fits and the data for the 
three cases with significantly diminished emission in their central regions.  
The panels are as in Figure \ref{results}, although note the slightly different 
bin sizes (30\,k$\lambda$) for the DoAr 44 visibility profile.  High resolution 
inset images are constructed for the WSB 60 disk, showing (to scale) the 
detailed structure of the continuum emission on the smallest spatial scales.  
Each inset is 1\farcs3 ($\sim$160\,AU) on a side, centered on the continuum 
peak in the main image (which is intentionally offset there for clarity).  
Contour levels in the inset are at intervals of 10\,mJy beam$^{-1}$.  Only a 
crude inner hole model is used here; more detailed examination of the emission 
morphologies and SEDs will be made elsewhere.  \label{trans}}
\end{figure}

\begin{figure}
\epsscale{0.5}
\plotone{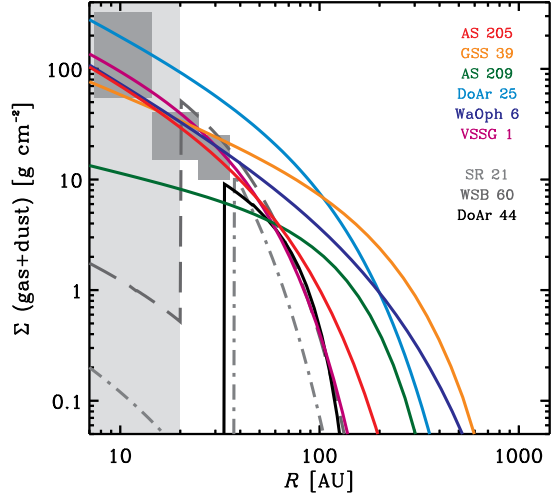}
\figcaption{Radial surface density (gas+dust) profiles derived for the sample 
disks, based on the parameters in Tables \ref{structure_table} and 
\ref{trans_table} and equations (4) and (5).  The three disks with central 
emission cavities are shown with grayscale curves (SR 21 as {\it dash-dot}, WSB 
60 as {\it long-dash}, and DoAr 44 as {\it solid}).  The dark gray rectangular 
regions mark the surface densities for Saturn, Uranus, and Neptune in the 
Minimum Mass Solar Nebula, reconstructed from current planet masses augmented 
to solar composition and smeared into annuli \citep[see][and \S 
5]{weidenschilling77}.  The sample disks are well-characterized by radial 
surface density gradients in the range $\gamma = 0.4$-1.0 with a median value 
$\gamma = 0.9$, and a range of masses and characteristic size scales.  The 
light gray shaded region marks the maximum resolution scale of the SMA 
observations.  \label{sigma}}
\end{figure}

\begin{figure}
\epsscale{0.5}
\plotone{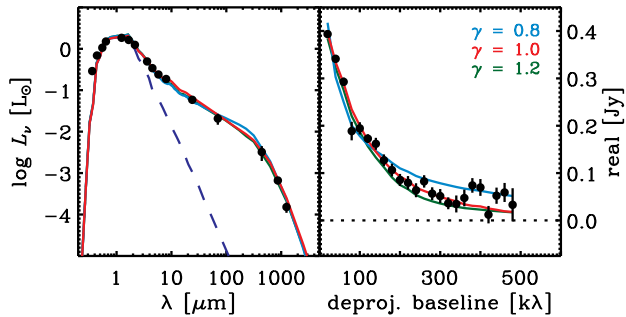}
\figcaption{Qualitative demonstration of the accuracy in the fitted $\gamma$
values, as discussed in \S 3.4.  The SED ({\it left}) and visibility profile 
({\it right}) are shown for the WaOph 6 disk, as in Figure \ref{results}.  The 
red curves again show the best-fit model ($\gamma = 1.0$; see Table 
\ref{structure_table}), while the blue and green curves show the models that 
best reproduce the data for $\gamma = 0.8$ and 1.2, respectively.  While all 
$\gamma$ values can reproduce the observed SED, those with departures of 
$\pm0.2$ from the best-fit $\gamma$ value show clear differences with the 
millimeter visibilities.  \label{gamma_unc}}
\end{figure}

\paragraph{AS 205 --}
Located in the northern outskirts of the main Oph clouds, AS 205 is a 
hierarchical triple system: the K5 primary (AS 205 A) lies 1\farcs3 northeast 
of a K7/M0 spectroscopic binary \citep[AS 205 B;][]{ghez93,prato03,eisner05}.  
Roughly 95\%\ of the 860\,$\mu$m emission from the system originates in a disk 
around the primary, which is the focus in this paper.   The remainder 
(50-60\,mJy) is associated with the AS 205 B binary (see Fig.~\ref{images}).  
Bright CO $J$=3$-$2 emission ($\sim$65\,Jy km s$^{-1}$) centered on the primary 
star shows a small velocity gradient along the disk major axis, indicative of a 
low inclination ($i = 25$\degr) viewing geometry (Fig.~\ref{moments}).  
However, spatio-kinematic asymmetries in the direction of the secondary are 
present in the CO channel maps, suggesting that some of the line emission is 
associated with AS 205 B.  Previous interferometric observations were unable to 
resolve the emission from the AS 205 A/B components \citep{aw07}.  

We constructed a SED for AS 205 A using component-resolved photometry from 
$\sim$0.5-12\,$\mu$m \citep{coku79,liu96,mccabe06}.  Composite 25 and 
60\,$\mu$m flux densities \citep{weaver92} were split based on a simple 
extrapolation of the resolved A and B photometry and ground-based spectra 
\citep{schegerer06} at shorter wavelengths.  Millimeter flux densities were 
determined by partitioning single-dish photometry \citep[350, 450, and 
1300\,$\mu$m;][]{aw07,aw07b} according to the resolved 860\,$\mu$m A/B flux 
ratio found here.  The resulting infrared SED is the brightest in the sample 
relative to the stellar photosphere, driving the model fits to infer a 
particularly extended vertical structure.  But given the complications 
associated with the small angular separation between the A and B components, a 
combined structure analysis of both disks would be beneficial in the future.

\paragraph{GSS 39 --}
This heavily reddened M0 star in the L1688 dark cloud hosts a bright, 
well-resolved millimeter continuum disk (Fig.~\ref{images}).  Resolved CO 
$J$=3$-$2 emission is found centered on GSS 39, with a velocity gradient 
oriented along the major axis of the continuum disk.  However, there is 
substantial contamination from cloud material that prohibits a more detailed 
analysis.  The high extinction ($A_V \approx 15$) produced by that cloud 
material forces us to base our stellar luminosity estimate on photometry in the 
near-infrared \citep[2MASS;][]{cutri03}, rather than at optical wavelengths.  
Given the non-negligible disk emission at $\lambda \ge 1$\,$\mu$m, there is 
considerable uncertainty on $L_{\ast}$ that propagates to the disk structure 
parameter estimates.  The remainder of the SED was collected from the {\it 
Spitzer} c2d Legacy project \citep{evans03}, ground-based infrared photometry 
\citep{lada84}, and 350-1200\,$\mu$m single-dish flux densities 
\citep{stanke06,aw07b}.  The GSS 39 disk is the only case our study has in 
common with the recent 1.3\,mm interferometric survey by \citet{isella09}.  
Despite using different modeling procedures, we find similar disk parameters 
within the quoted uncertainties.

\paragraph{AS 209 --} 
This young K5 star, relatively isolated from the main Ophiuchus clouds, was 
found to harbor a Keplerian molecular gas disk by \citet{koerner95}.  That 
structure is confirmed here (Figs.~\ref{moments} and \ref{AS209_CO}), where the 
spatially resolved CO $J$=3$-$2 line emission ($\sim$16.5\,Jy km s$^{-1}$) has 
a clear rotation signature consistent with a disk inclined $\sim$40\degr\ from 
face-on.  While broadband SED information for AS 209 is sparse, we adopt 
averaged optical data from variability studies \citep{herbst94,grankin07}, 
ground-based and {\it IRAS} infrared photometry 
\citep{hamann92,weaver92,cutri03}, and single-dish millimeter flux densities 
\citep{andre94,aw07b}.  A publicly available $\sim$5-30\,$\mu$m {\it Spitzer} 
IRS spectrum from the c2d Legacy project \citep{evans03} is dominated by broad 
silicate emission features, and therefore not used in our modeling.  The 
absence of useable $\sim$5-60\,$\mu$m SED data means that the inferred vertical 
structure is particularly uncertain in this case.  Moreover, it implies that 
the millimeter visiblities play a comparatively larger role in the parameter 
estimates.  Perhaps the lack of infrared SED data contributes to 
the AS 209 disk having the lowest $\gamma$ value (0.4) in the sample.

\paragraph{DoAr 25 --}
A K5 star in the L1688 dark cloud, DoAr 25 hosts a resolved optical/infrared 
scattered light disk \citep[][private communication]{stapelfeldt08} that is 
also particularly bright at millimeter wavelengths.  Despite these substantial 
disk signatures, the standard hallmarks of accretion for the DoAr 25 disk are 
comparatively meager: estimates of the mass accretion rate range from low 
\citep[$3\times10^{-9}$\,M$_{\odot}$ yr$^{-1}$;][]{greene96,luhman99} to 
negligible \citep[$<2\times10^{-10}$\,M$_{\odot}$ yr$^{-1}$;][]{natta06} 
values.  The SED also shows a relatively small excess over the stellar 
photosphere from $\sim$1-15\,$\mu$m (Fig.~\ref{results}) compared to typical T 
Tauri disks, although a more substantial excess is present at longer 
wavelengths.  We constructed the DoAr 25 SED from optical data 
\citep{vrba93,wilking05}, the 2MASS database \citep{cutri03}, {\it Spitzer} 
observations \citep{evans03,padgett08}, and single-dish millimeter photometry 
\citep{andre94,dent98,aw07b}.

In an effort to model these same SMA data while accounting for the small 
infrared excess and low accretion rate, \citet{andrews08} suggested that the 
DoAr 25 disk has a shallow density distribution, $\Sigma \propto R^{-0.34}$.  
We find a steeper inner disk $\Sigma$ profile in our fits ($\Sigma \propto 
R^{-0.9}$ when $R \lesssim 80$\,AU), which utilize different assumptions for 
some key aspects of the modeling process: (1) a smaller distance (125\,pc, 
compared to 145\,pc) shortens spatial scales and makes radial profiles steeper; 
(2) a smaller $L_{\ast}$ (0.8\,L$_{\odot}$, compared to 1.3\,L$_{\odot}$) 
decreases disk temperatures and leads us to infer different vertical structure 
parameters; and (3) a more appropriate treatment of the visibilities in the 
fitting process (i.e., using each visibility point, rather than the binned 
visibility profile) places less emphasis on a detailed match to the infrared 
SED (see \S 3.4).  Because of these differences, we find a best-fit model with 
a larger flaring angle ($\psi = 0.15$, compared to 0.11), leading to a flatter 
midplane $T$ profile that requires a comparatively steeper $\Sigma$ profile to 
reproduce the observed visibilities (see \S 4.1.4).  While we find a lower 
$\chi^2$ value than for the \citet{andrews08} results, we note that $\Sigma$ 
profiles with gradients as low as $\gamma \approx 0.5$ are also able to 
reproduce the data fairly well.  A more robust conclusion could be made in a 
future modeling study that incorporates the scattered light morphology and a 
more detailed infrared spectrum to help independently constrain the vertical 
structure of the disk.

\paragraph{WaOph 6 --}
This K6 star is located north of the main Oph clouds, close to AS 209.  In 
addition to a compact millimeter continuum emission morphology, the WaOph 6 
disk exhibits CO $J$=3$-$2 line emission ($\sim$13.5\,Jy km s$^{-1}$) with a 
clear velocity gradient along the disk major axis.  While the spatio-kinematics 
of that CO emission in the channel maps is consistent with a Keplerian rotation 
pattern inclined $\sim$40\degr\ from face-on, there is some cloud contamination 
near the systemic velocity.  The SED used here is composed of the 
optical/infrared measurements by \citet{eisner05}, {\it Spitzer} data from 
\citet{padgett06}, and single-dish millimeter photometry 
\citep{andre94,aw07b}.  The infrared SED has a steep spectral slope 
(Fig.~\ref{results}), consistent with a flat disk geometry where dust grains 
have largely settled toward the disk midplane.  This vertical structure results 
in a steep midplane temperature profile (Fig.~\ref{Tmid}), with a particularly 
cold outer disk.  Despite the large $R_c$ value inferred for the WaOph 6 disk, 
those low temperatures at large radii lead to the observed compact emission 
morphology.

\paragraph{VSSG 1 --}
The disk around this heavily extinguished M0 star in the L1688 cloud has the 
most compact, centrally concentrated millimeter emission in the sample.  Bright 
CO $J$=3$-$2 emission with significant cloud contamination is found around the 
star in two distinct LSR velocity intervals centered at $-$3.0 and $-$0.4\,km 
s$^{-1}$.  It is unclear if any of that emission is associated with the disk.  
As with the GSS 39 disk, high extinctions force us to estimate an uncertain 
$L_{\ast}$ in the near-infrared \citep{cutri03}.  The ambiguity in that 
luminosity estimate translates directly to our estimates of disk structure 
parameters.  The remainder of the SED was compiled from the c2d {\it Spitzer} 
database \citep{evans03} and single-dish millimeter photometry data 
\citep{stanke06,aw07b}.  The infrared SED is similar to WaOph 6, with a steep 
spectral gradient indicative of a relatively flat geometry.  Given the compact 
continuum emission, it is not surprising that we find a small radius ($R_c = 
33$\,AU).

\paragraph{SR 21 --}
A large central cavity in the disk around this G3 star was first inferred from 
the broadband SED \citep{brown07}, and then confirmed with the archival (V 
configuration) SMA data presented here \citep{pontoppidan08,brown09}.  The SED 
shown in Figure \ref{trans} was compiled from optical \citep{chini81,vrba93} 
and near-infrared photometry \citep{cutri03}, the {\it Spitzer} c2d Legacy 
database \citep{evans03}, older {\it IRAS} measurements \citep{weaver92}, and 
millimeter photometry from 350\,$\mu$m to 2.7\,mm 
\citep{andre94,aw07b,patience08}.  Our new lower resolution (S configuration) 
SMA observations do not show any clear evidence for CO $J$=3$-$2 emission from 
the SR 21 disk, although cloud contamination may obscure a weak signal.  On the 
contrary, \citet{pontoppidan08} find infrared emission lines from molecular gas 
inside the SR 21 disk cavity.  With a simplistic model, we place the inner edge 
of that cavity at $R_{\rm cav} \approx 37$\,AU, further than the 18\,AU 
inferred by \citet{brown07} from the SED alone.  But considering the major 
differences in model assumptions (particularly the stellar properties), the 
discrepancy is likely not significant.  A more in-depth modeling analysis of 
the SED and resolved continuum data together will be performed elsewhere 
\citep{brown09}, and should provide more robust constraints on the cavity 
dimensions.

\paragraph{WSB 60 --}
With a spectral type of M4, this cool star in the L1688 cloud is significantly
less massive than the rest of the sample.  Nevertheless, it harbors one of the
brightest millimeter disks in the Oph star-forming region, with a remarkable 
low-density cavity noted on scales near the SMA angular resolution limit 
($R_{\rm cav} \approx 20$\,AU).  The SED used here was compiled from the 
optical survey by \citet{wilking05}, the 2MASS database \citep{cutri03}, {\it 
Spitzer} photometry \citep{evans03,padgett08}, and integrated millimeter flux 
densities \citep[350-1300\,$\mu$m;][]{aw07,aw07b}.  The cavity imaged in the 
inset of Fig.~\ref{trans} can be confirmed with a close examination of the 
visibility profile, which shows that the real part of the visibility fluxes lie 
below zero on deprojected baselines longer than $\sim$350\,k$\lambda$.  With 
such a low stellar mass and a large central cavity, this source certainly 
merits a follow-up modeling study with a more complete SED dataset.

\paragraph{DoAr 44 --}
This young K3 star in the L1688 cloud appears to be a typical classical T Tauri 
star, and was selected for this survey only because of its brighter than 
average millimeter emission.  The broadband SED is typical of a normal 
continuous disk, constructed here from optical data \citep{herbst94}, the 2MASS 
database \citep{cutri03}, ground-based images \citep{greene94}, {\it Spitzer} 
photometry \citep{evans03,padgett08}, and single-dish millimeter flux densities 
\citep{nurnberger98,aw07b}.  Nevertheless, the observed double-peaked 
morphology of the millimeter emission for the DoAr 44 disk (Fig.~\ref{images}) 
is a clear signature of an inclined ring, with a lack of emission 
in the inner disk.  Although we make a preliminary estimate for the inner edge 
of that disk cavity at $R_{\rm cav} \approx 33$\,AU, there is substantial 
uncertainty due to the poor fit of the infrared SED.  A relatively bright 
H$\alpha$ line \citep[42\AA;][]{guenther07} and substantial infrared excesses 
confirm that there is a substantial amount of material inside this millimeter 
emission cavity.  A more sophisticated effort to model the inner region of the 
DoAr 44 disk using the SED, SMA visibilities, and the {\it Spitzer} IRS 
spectrum will be presented elsewhere.

\section{Discussion}

\subsection{Viscous Disk Evolution}

For most of its lifetime, the structural evolution of a disk will be determined
by the viscous and gravitational interactions that govern the accretion 
process.  The Keplerian rotation and anomalous turbulent viscosities of the 
disk material combine to drive a net mass flow inwards to smaller radii 
\citep{lyndenbell74}, where magnetic fields can channel it onto the stellar 
surface \citep[e.g.,][]{konigl91,shu94}.  Meanwhile, to balance the angular 
momentum lost in that process, some disk material is transported out to larger 
radii \citep[e.g.,][]{pringle81}.  These coupled effects act to decrease 
densities at smaller radii as the disk expands: the disk structure literally 
spreads itself thin over time.  Empirical constraints on this viscous mode of 
disk structure evolution are critical for developing improved models of planet 
formation and various disk dissipation mechanisms.

For the case where disk viscosities are static (constant with time) and 
spatially distributed like a power-law, $\nu \propto R^{\gamma}$, 
\citet{lyndenbell74} derived an analytic similarity solution for the viscous 
evolution of the disk surface density profile.  The form of that similarity 
solution \citep[see][and the Appendix]{hartmann98} is identical to the one 
adopted in equation (4) when
\begin{eqnarray}
R_c & = & R_1 {\cal T}^{1/(2-\gamma)} \\
M_d & = & M_{d,0} {\cal T}^{-1/2(2-\gamma)},
\end{eqnarray}
where $R_1$ is a scaling radius, ${\cal T}$ is a dimensionless time parameter, 
and $M_{d,0}$ is the initial disk mass.  The time parameter ${\cal T} = 1 + 
t/t_s$, where $t$ is the time since viscous evolution began and $t_s$ is the 
viscous timescale.  In essence, ${\cal T}$ tracks how many viscous timescales 
have elapsed.  In its initial configuration ($t = 0$; ${\cal T} = 1$), 
$\sim$60\%\ of the disk mass was concentrated inside the radius $R_1$.  

\citet{hartmann98} explored the prominent role that the surface 
density/viscosity gradient ($\gamma$) plays in the evolution of observable disk 
structure parameters.  For a given set of initial conditions, they showed that 
a larger $\gamma$ value leads to accelerated viscous spreading (expansion to 
conserve angular momentum).  That relationship is encapsulated in equation (7), 
where the characteristic radius ($R_c$) determined in our model fits is an 
empirical measure of how much the disk has expanded since its initial 
configuration ($R_c = R_1$ when ${\cal T} = 1$).  For the typical value $\gamma 
= 0.9$ found in this sample, $R_c$ increases roughly linearly with ${\cal T}$.  
Because it is not clear when the evolutionary clock for a given disk was 
started (${\cal T}$ is unknown), a precise calibration of the initial 
conditions \{$R_1$, $M_{d,0}$\} for these viscous models is a challenge.  
Associating ${\cal T}$ with stellar ages can be problematic, as this simple 
viscous evolution behavior may not have been applicable for a significant 
fraction of that time.  Figure \ref{accretion} illustrates how these initial 
conditions depend on the unknown value of ${\cal T}$, based on the current disk 
structures derived in \S 4 (see equations [7-8]).  Note that these plots do not 
track the evolution of \{$R_1$, $M_{d,0}$\}: the initial conditions {\it do not 
change with time}.  Rather, they mark what their appropriate values would be 
for a given value of ${\cal T}$.  For example, if 10 viscous timescales have 
elapsed since the WaOph 6 disk began its evolution (${\cal T} = 11$), then the 
current disk structures imply that the initial disk mass was $M_{d,0} \approx 
0.27$\,M$_{\odot}$, and 60\%\ of that mass (0.16\,M$_{\odot}$) was initially
contained inside $R_1 \approx 12$\,AU.  Similar examples for each disk can be 
made from the initial conditions in Table \ref{viscous_table}, where we have 
listed what the \{$R_1$, $M_{d,0}$\} values would be if 1 or 10 viscous 
timescales have elapsed (${\cal T} = 2$ and 11).

Aside from providing some limited insight on the initial conditions, our 
estimates of \{$\gamma$, $R_c$\} help characterize another key radius  
\begin{equation}
R_t = R_c \left[\frac{1}{2(2-\gamma)}\right]^{1/(2-\gamma)}
\end{equation}
that marks where the mass flow through the disk changes direction (to conserve 
angular momentum); material moves in toward the central star for $R < R_t$ and 
out to larger disk radii for $R > R_t$.  In this sample, we find $R_t$ values 
of $\sim$10-100\,AU (see Table \ref{viscous_table}; note that $R_t = R_c/2$ 
when $\gamma = 1$).  Aside from these constraints on $R_t$, the only 
observational diagnostics of the mass flow in disks are the stellar accretion 
rates, $\dot{M}_{\ast}$.  \citet{hartmann98} demonstrated that those accretion 
rates would decay more rapidly with ${\cal T}$ for larger $\gamma$ values.  
They used a power-law fit to the empirical $\dot{M}_{\ast}$-age relation for T 
Tauri disks to infer that $\gamma = 1$ (or larger), in excellent agreement with 
the results from our parametric structure fits \citep[see also][]{calvet00}.

Only a few observational studies have attempted to constrain disk surface 
densities in the context of these viscous evolution models.  \citet{kitamura02} 
fitted the SEDs and 2\,mm images ($\sim$1\arcsec\ resolution) for 13 disks with 
a 1-D structure model.  They found a range of $\gamma$ values from 0.0-0.8, and 
comparable $R_c$ values that appeared to be anticorrelated with the H$\alpha$ 
line luminosities.  \citet{hughes08} used a simple (vertically isothermal) 2-D 
structure model to reproduce both the continuum and CO line visibilities 
($\sim$1-2\arcsec\ resolution) for 4 disks with $\gamma = 0.7$-1.1.  And most 
recently, \citet{isella09} fitted the 1.3\,mm visibilities ($\sim$0\farcs7 
resolution) for 14 disks with a ``two-layer" structure model 
\citep[cf.,][]{chiang97}.  Their results suggested a wide range of $\gamma$ 
values, from $-$0.5 to 0.8, and characteristic radii similar to those presented 
here.  \citet{isella09} also argue that their $R_t$ values are correlated with 
stellar age.  Spatial resolution mismatches, distinct modeling styles, and 
deviating opacity prescriptions all play some role in explaining the 
differences in the $\gamma$ values inferred here (and by Hughes et al.)~and 
those at the low end of the ranges found by \citet{kitamura02} and 
\citet{isella09}.  We do not find any convincing evidence for correlations 
between the disk structures and stellar properties or accretion diagnostics, 
but this may be the result of limitations in the sample size and properties.

Various theoretical perspectives have argued for $\gamma$ values similar to 
those found with our parametric structure modeling.  \citet{hartmann98} pointed 
out that $\gamma \approx 1$ would be expected for an irradiated disk in 
vertical hydrostatic equilibrium if the turbulent viscosity parameter, 
$\alpha$, is roughly constant with radius (see below).  In a new approach that 
treats the viscous evolution of magnetized disk material, \citet{shu07} 
predicted steep surface density profiles with $\Sigma \propto R^{-3/4}$.  A 
sequence of sophisticated numerical modeling studies that self-consistently 
track both the formation and subsequent evolution of disks have suggested that 
larger values near $\gamma \approx 1.5$ are more appropriate \citep[][but see 
Hueso \& Guillot 2005]{vorobyov07,vorobyov08}.  The $\Sigma$ profiles in those 
models correspond to gravitationally unstable, non-axisymmetric disk structures 
with significantly more mass than has been estimated from observations.  
However, \citet{vorobyov09} demonstrate that those instabilities and azimuthal 
asymmetries would quickly damp out when turbulent viscosities are included in 
the simulations, producing $\Sigma$ profiles ($\gamma \approx 0.8$), disk 
masses, and accretion rates that are more in line with those estimated here and 
elsewhere.

While observational constraints on surface density profiles help characterize 
how material is redistributed over time within these disks, the pace of that 
evolution is set by the disk viscosities.  The physical origin of those 
viscosities remains uncertain, although various mechanisms that could drive 
large-scale turbulent motions in these disks have been suggested as 
possibilities.  \citet{shakura73} proposed a simple prescription for turbulent 
viscosities, $\nu = \alpha c_s H$, where $c_s$ is the sound speed, $H$ is the 
scale height, and $\alpha$ is a key parameter that characterizes the efficiency 
of angular momentum transport.  Incorporating that relation into our 
formulation for the structure of viscous disks, we can solve for the viscosity 
parameter
\begin{equation}
\alpha \approx \frac{\dot{M}_{\ast}}{3 \pi \Sigma_c} \left(\frac{R}{R_c}\right)^{\gamma} \frac{1}{c_s H}
\end{equation}
where $\Sigma_c$ is defined in equation (5) (see the Appendix).  For the disk 
structures and accretion rates in this sample, we find the wide distribution of 
$\alpha$ values shown in Figure \ref{alphas} and listed in Table 
\ref{viscous_table}, ranging from 0.0005 to 0.08.  These calculations use the 
$\dot{M}_{\ast}$ values in Table \ref{viscous_table}, which were derived 
directly from emission line fluxes (or $U$-band excesses) in the literature 
using the standard data$-$accretion luminosity conversions 
\citep{muzerolle98,calvet04} to force consistency with our adopted stellar 
properties.  The histogram in Figure \ref{alphas} technically corresponds to 
$\alpha$ measured at $R = 10$\,AU, but we show in the Appendix that these 
values have only a weak radial dependence. The $\alpha$ values inferred from 
the disk structures are in good agreement with those found in numerical 
magnetohydrodynamics simulations, where the magnetorotational instability 
\citep{balbus91} is the mechanism responsible for generating the anomalous 
turbulent viscosities \citep[e.g.,][]{hawley95,stone96,fleming03,fromang07}.

\subsection{Planet Formation Potential}

Although the surface density profiles for this sample are in good agreement 
with viscous disk models, they are different than the common prescription for 
the primordial solar disk (the MMSN).  As with the model fits in \S 4, there 
are considerable uncertainties in constructing a $\Sigma$ profile for the MMSN: 
the abundances of key elements in current planet compositions, the arbitrary 
assignment of annular bins to smear out augmented masses, and the implicit 
assumption that the nebular material was stationary throughout its evolution, 
to name a few \citep[see][]{cameron88,desch07,crida09}.  But if the zones 
around Mercury and Mars are ignored due to the high likelihood of their 
dynamical depletion, the {\it in situ} augmented planetary densities can be fit 
reasonably well with a $\Sigma \propto R^{-1.5}$ profile 
\citep{weidenschilling77,hayashi81}.  A steeper $\Sigma \propto R^{-2.2}$ 
profile has been argued for by \citet{desch07}, who uses modified planetary 
accretion zones motivated by dynamical simulations of the early solar system.  
A similar $\Sigma \propto R^{-2}$ profile was determined from the combined 
properties of 12 extrasolar multi-planet systems \citep{kuchner04}.  As an 
alternative, \citet{davis05} has suggested a viscous disk model with a flatter 
inner disk $\Sigma$ profile corresponding to $\gamma \approx 0.5$. 

A direct comparison of the density structures for the sample disks and the MMSN 
was shown in Figure \ref{sigma}.  Alternatively, Figure \ref{Mcum} compares how 
the encircled mass varies with radius in the disks and MMSN.  The latter is 
defined as the cumulative mass interior to a given radius, $M$($<$$R$) $= \int 
\Sigma \,\, 2 \pi R \,\, dR$ from $R_{\rm in}$ to $R$.  Unlike the $\Sigma$ 
profile, the MMSN encircled masses do not rely on the arbitrary widths and 
locations of annular bins.  The error bars reflect the uncertainties in the 
augmented $M$($<$$R$) values for Saturn, Uranus, and Neptune in the MMSN and 
those planets interior to them \citep[particularly 
Jupiter;][]{weidenschilling77}.  The dark gray region shows the encircled 
masses implied by the standard $\Sigma \propto R^{-1.5}$ MMSN profiles of 
\citet[][{\it lower bound}]{hayashi81} and \citet[][{\it higher 
bound}]{weidenschilling77}.  Despite their flatter $\Sigma$ profiles, Figures
\ref{sigma} and \ref{Mcum} illustrate that the mass distributions for the 
sample disks are comparable to the MMSN values, within the uncertainties.  
Particularly good agreement is noted in the resolved Uranus-Neptune region 
(with the exception of the AS 209 disk).  The sample disks have $\Sigma = 
30$-100\,g cm$^{-2}$ at 19\,AU, 15-60\,g cm$^{-2}$ at 30\,AU, and contain 
0.02-0.05\,M$_{\odot}$ of material inside 40\,AU, while the standard MMSN 
models have 15-40\,g cm$^{-2}$ at 19\,AU, 10-25\,g cm$^{-2}$ at 30\,AU, and 
0.01-0.04\,M$_{\odot}$ inside 40\,AU.

However, we should emphasize that angular resolution limitations prevent a 
direct characterization of the mass contents in these disks on radial scales 
smaller than $\sim$20\,AU.  In particular, the data cannot differentiate the 
assumed $\Sigma$ formulation (\S 3.1) from alternative density models at 
smaller radii (e.g., cases where the viscosity may not be a simple power law).  
Therefore, since the constraints on $\Sigma$ are made from the millimeter 
emission on much larger spatial scales, extrapolations of the density profiles 
into the inner giant planet region (i.e., in to 5\,AU) remain speculative.  We 
do find a notable difference between the sample disks and MMSN mass contents at 
large radii.  The solar system has a relatively sharp edge at $\sim$40\,AU 
\citep{luu02}, while the sample disks have substantial mass reservoirs on 
larger scales.  The MMSN disk could have been externally truncated, from an 
encounter with a passing star \citep{ida00} or an intense radiation field 
\citep{johnstone98} in a now-dispered massive star cluster.  

In planet formation models based on the canonical ``core accretion" scenario, a 
scaled-up MMSN disk can produce giant planets on timescales of $\sim$3-8\,Myr 
\citep{pollack96,inaba03,hubickyj05,lissauer07}, roughly consistent with the 
upper bound on disk lifetimes \citep[e.g.,][]{haisch01}.  But \citet{alibert05} 
demonstrated that those formation timescales are reduced to only $\sim$1\,Myr 
if nominal viscous evolution and migration effects are included in such 
calculations.  Moreover, the ability to form planets in the core accretion 
scenario does not appear to be significantly different for $\Sigma$ profiles 
that are flatter than the MMSN (like the $\gamma \approx 0.9$ profiles inferred 
here), provided that there is sufficient mass in the inner disk 
\citep{wetherill96,chambers02,raymond05}.  Alternatively, the ``disk 
instability" scenario can potentially form giant protoplanets via the 
fragmentation of a scaled-up MMSN disk on timescales as short as $\sim$1000\,yr 
\citep{cameron78,boss97,mayer07,durisen07}.  But both of these planet formation 
scenarios require surface densities $\ge$2-5$\times$ larger than those found 
for the sample disks or the MMSN, either to accomodate key timescale 
constraints (core accretion) or to maintain a gravitationally unstable 
structure (disk instability).  Assuming that those requirements are accurate, 
it is worthwhile to consider ways to reconcile this $\Sigma$ discrepancy. 

Perhaps the most straightforward way to do that is associated with the disk 
opacities ($\kappa$).  Because the millimeter emission from a disk is related 
to the product $\kappa \Sigma$, changes to the dust opacities would be 
reflected in the inferred surface densities.  The opacities regulate the 
conversion of surface brightnesses to surface densities, with an exchange rate 
set by the properties and size distribution of disk solids 
\citep[e.g.,][]{beckwith91}.  The disk structures we mean to compare with 
planet formation models were determined for a dust grain size distribution 
$n(a) \propto a^{-3.5}$ up to a maximum $a = 1$\,mm (see \S 3.2), and assuming 
that the dust tracks 1\%\ of the total mass budget (the remainder is molecular 
gas).  But different size distributions tend to produce {\it lower} millimeter 
opacities, which could in turn lead us to infer {\it higher} surface densities 
from the same data \citep{miyake93,dalessio01,draine06}.  For example, 
extending the same size distribution out to 1\,cm particles decreases the 
870\,$\mu$m opacity by a factor of $\sim$2.6 \citep{dalessio01}, and would 
scale our $\Sigma$ estimates up to be in line with the planet formation model 
requirements without dramatically affecting the estimates of other structure 
parameters.  Such modest adjustments to the particle population are not 
unreasonable considering the results of grain growth simulations 
\citep{weidenschilling97,dullemond05,garaud07} and the emission signatures of 
centimeter-scale grains in similar disks \citep{natta04,wilner05,rodmann06}.  
Indeed, the core accretion models assume that a large population of 
planetesimals is already present in these disks, implying that the dust we 
observe is merely collisional debris \citep[e.g.,][]{kenyon09}.  

Therefore, with only small modifications to the model assumptions we can 
conclude that the giant planet formation process is indeed feasible for the 
disks in this sample.  In fact, planet formation may have already started in 
the disks around SR 21, WSB 60, and DoAr 44.  The ring-like morphology of the 
millimeter emission from those disks could be the signature of a low-density 
cavity on size scales comparable to the solar system.  While a variety of 
mechanisms could produce this kind of structure 
\citep[see][]{dalessio05,najita07}, perhaps the most compelling is the presence 
of a giant planet just inside the ring edge.  A sufficiently massive planet can 
open a gap in the disk structure, significantly decreasing (or even halting) 
the mass flow through to the inner disk 
\citep[e.g.,][]{lin86,bryden99,lubow99,lubow06,varniere06}.  Consequently, 
accretion depletes the inner disk densities and diminishes the millimeter 
emission interior to the planet.  But why might planets have formed in these 
three disks, and not the other six in this sample?  Perhaps they formed planets 
early, when densities were high enough to facilitate the gravitational 
instability scenario.  Or maybe they have simply had more time than the others 
to form a giant planet with the core accretion scenario: we find stellar ages 
of 3-7\,Myr for these cases (Table \ref{stars_table}).  While individual 
stellar age estimates are plagued with uncertainty \citep{hillenbrand04}, it is 
interesting to note that the host stars for all of the confirmed ``transition" 
disks $-$ with resolved images of their cavities $-$ seem to be older than a 
few Myr \citep[e.g.,][]{pietu06,brown08,brown09,hughes07,hughes09}.

\section{Summary}

We have presented the results of a high angular resolution (0\farcs3 $\approx 
40$\,AU) survey of the 870\,$\mu$m continuum emission from 9 young 
circumstellar dust disks in the Ophiuchus star-forming region.  Those 
observations were used to model some of the key size scales and physical 
conditions in these disks, based on two-dimensional Monte Carlo radiative 
transfer calculations.  We find the following:
\begin{enumerate}
\item For a parametric surface density profile $\Sigma \propto 
(R/R_c)^{-\gamma} \exp{[-(R/R_c)^{2-\gamma}]}$, the radial $\Sigma$ gradients 
derived for the sample disks lie in a narrow range around a median value 
$\gamma = 0.9$ (from $\gamma = 0.4$-1.0).  Therefore, the shapes of the surface 
density profiles for these disks are similar, varying roughly inversely with 
radius in the inner disk ($\Sigma \propto 1/R$ when $R \lesssim R_c$) and 
smoothly merging into a steeper exponential decrease at larger radii ($\Sigma 
\propto 1/e^R$ when $R \gtrsim R_c$).  The characteristic radii that mark that 
transition in the $\Sigma$ profile lie in the range $R_c \approx 20$-200\,AU.  
Because these disks were selected to have bright millimeter emission, the 
sample is biased toward high disk masses, $M_d \approx 
0.005$-0.14\,M$_{\odot}$.  A variety of vertical structures are also 
determined, from significantly flared to nearly flat geometries.

\item These structure constraints are used to help characterize the viscous 
properties of the disk material.  The inferred $\gamma$ values are lower than 
expected for gravitationally unstable disk models \citep[$\gamma \approx 
1.5$; e.g.,][]{vorobyov09}, but are generally consistent with simple models 
that incorporate a prescription for turbulent disk viscosities \citep[$\gamma 
\approx 1$; e.g.,][]{hartmann98}.  Combining these disk structures with 
accretion rate measurements and a simple viscous disk model, we estimate 
turbulent viscosity parameters in the range $\alpha \approx 0.0005$-0.08.  

\item Disk densities in the outer giant planet region ($R \approx 20$-40\,AU) 
are comparable to the expected values for the reconstructed density structure 
of the primordial solar disk (MMSN), despite their generally less steep 
$\Sigma$ profiles in the inner disk.  But unlike the typical models of the 
MMSN, the sample disks have substantial mass reservoirs at large distances from 
the star ($R > 40$\,AU).  Although current planet formation models need at 
least 2-5$\times$ higher densities to operate efficiently, we argue that small 
modifications to our assumed opacities that are consistent with modest dust 
grain growth can easily accomodate those model requirements.  

\item Three of the sample disks (SR 21, WSB 60, and DoAr 44) exhibit resolved 
ring morphologies, where the continuum emission is significantly diminished 
inside a radius of $\sim$20-40\,AU.  A variety of mechanisms could potentially 
decrease densities or opacities on these scales in the inner disk, and should 
be explored in more detail for these cases.  One interesting possibility is 
that these disks have already produced massive planets just inside their ring 
edges.
\end{enumerate}

\acknowledgments We thank Michiel Hogerheijde for kindly providing access to 
the 2-D version of the {\tt RATRAN} code package, Paola D'Alessio for 
calculating various opacity spectra on our behalf, Andrea Isella for sharing 
results with us prior to publication, and an anonymous referee for a thoughtful 
and very useful review.  The SMA is a joint project between the Smithsonian 
Astrophysical Observatory and the Academia Sinica Institute of Astronomy and 
Astrophysics and is funded by the Smithsonian Institution and the Academia 
Sinica.  Support for this work was provided by NASA through Hubble Fellowship 
grant HF-01203.01-A awarded by the Space Telescope Science Institute, which is 
operated by the Association of Universities for Research in Astronomy, Inc., 
for NASA, under contract NAS 5-26555.  D.~J.~W.~acknowledges support from NASA 
Origins Grant NNG05GI81G.

\appendix

\section{Accretion Disks}

This appendix is intended to facilitate a clear comparison between our 
parametric structure models (\S 3) and the properties of viscously evolving 
accretion disks \citep[as presented by][]{hartmann98}.  The evolution of the 
density structure of a viscous disk in Keplerian rotation around a central star 
is characterized with the partial differential equation
\begin{equation}
\frac{\partial \Sigma}{\partial t} = \frac{3}{R} \frac{\partial}{\partial R} \left[ R^{1/2} \frac{\partial}{\partial R} (R^{1/2} \nu \Sigma) \right],
\end{equation}
where $\nu$ denotes the viscosity \citep{lyndenbell74,pringle81}.  If the 
viscosities do not change with time and are spatially distributed as a 
power-law, $\nu \propto R^{\gamma}$, there is an analytic similarity solution 
to (A1) with the form
\begin{equation}
\Sigma = \frac{{\cal C}}{3 \pi \nu_1} \left( \frac{R}{R_1} \right)^{-\gamma} {\cal T}^{-(5/2-\gamma)/(2-\gamma)} \exp{\left[- \left( \frac{R}{R_1} \right)^{2-\gamma} {\cal T}^{-1} \right]},
\end{equation}
where $R_1$ is a scaling radius, $\nu_1 = \nu(R_1)$, and ${\cal T}$ is a 
dimensionless time parameter such that ${\cal T} = 1+t/t_s$, where $t_s$ is the 
viscous timescale.  Note that ${\cal T}-1$ is equivalent to the number of 
viscous timescales that have elapsed since the evolution began ($t = 0$).  The 
constant ${\cal C}$ can be recast in terms of an initial disk mass, $M_{d,0}$, 
by integrating (A2) over the disk area at $t = 0$ (${\cal T} = 1$)
\begin{equation}
{\cal C} = \frac{3 (2-\gamma) \nu_1}{2 R_1^2} M_{d,0}
\end{equation}
leading to a simplified form for the surface density profile in (A2)
\begin{equation}
\Sigma = (2-\gamma) \left( \frac{M_{d,0}}{2 \pi R_1^2} \right) {\cal T}^{-(5/2-\gamma)/(2-\gamma)} \left( \frac{R}{R_1} \right)^{-\gamma} \exp{\left[- \left( \frac{R}{R_1} \right)^{2-\gamma} {\cal T}^{-1} \right]}.
\end{equation}
This $\Sigma$ profile resembles the parametric form used in our models
\begin{equation}
\Sigma = (2-\gamma) \left( \frac{M_d}{2 \pi R_c^2} \right) \left( \frac{R}{R_c} \right)^{-\gamma} \exp{\left[-\left(\frac{R}{R_c}\right)^{2-\gamma}\right]}
\end{equation}
where equations (4) and (5) in \S 3.1 have been combined to ease the 
comparison.  The radial behavior of the exponential surface density tapers in 
(A4) and (A5) are identical if
\begin{equation}
R_c = R_1 {\cal T}^{1/(2-\gamma)}
\end{equation}
as described in equation (7) of \S 5.1.  Inserting that relation back into (A4) 
and comparing with (A5) confirms the evolutionary behavior of the disk mass 
derived by \citet[][their equation 26]{hartmann98} and noted in equation (8) of 
\S 5.1,
\begin{equation}
M_d = M_{d,0} {\cal T}^{-1/2(2-\gamma)}.
\end{equation}

Some additional insight can be inferred from the behavior of the mass flow rate
\begin{equation}
\dot{M} = {\cal C} {\cal T}^{-(5/2-\gamma)/(2-\gamma)} \left[1 - \frac{2(2-\gamma)}{{\cal T}}\left(\frac{R}{R_1}\right)^{2-\gamma} \right] \exp{\left[-\left(\frac{R}{R_1}\right)^{2-\gamma} {\cal T}^{-1}\right]}
\end{equation}
for these viscous disk models \citep[][their equation 21]{hartmann98}.  To 
conserve angular momentum, the direction of the mass flow (i.e., the sign of 
$\dot{M}$) changes at a radius
\begin{equation}
R_t = R_1 \left[ \frac{{\cal T}}{2(2-\gamma)}\right]^{1/(2-\gamma)} = R_c \left[\frac{1}{2(2-\gamma)}\right]^{1/(2-\gamma)}
\end{equation}
such that the bulk flow is in toward the star ($\dot{M} > 0$) when $R < R_t$, 
and outwards to larger radii in the disk ($\dot{M} < 0$) when $R > R_t$.  The 
second relation in (A9), obtained using (A6), demonstrates that the model 
parameters \{$\gamma$, $R_c$\} provide important constraints on the mass flow 
in the disk.  The only other available mass flow diagnostics are the accretion 
rates onto the star, $\dot{M}_{\ast}$.  Substituting in (A3), (A6), and (A7), 
and recognizing that $\dot{M}_{\ast}$ corresponds to the value of $\dot{M}$ at 
very small radii ($R \ll R_c$), (A8) can be written in the form
\begin{equation}
\dot{M}_{\ast} \approx \frac{3(2-\gamma)}{2} \nu_1 \left(\frac{M_d}{R_c^2}\right) {\cal T}^{\gamma/(2-\gamma)}.
\end{equation}
This relation suggests that some of the key disk structure parameters that can 
be constrained observationally, \{$M_d$, $\gamma$, $R_c$, $\dot{M}_{\ast}$\}, 
are potential probes of the disk viscosities (i.e., $\nu_1$). 

In a typical parameterization for accretion disks \citep{shakura73}, the 
viscosity is proportional to the product of the sound speed ($c_s$) and scale 
height ($H$)
\begin{equation}
\nu = \alpha c_s H = \nu_1 \left(\frac{R}{R_1}\right)^{\gamma}
\end{equation}
with the proportionality constant $\alpha$ considered a diagnostic for the 
efficiency of angular momentum transport.  Here, the second equality is the 
definition of the viscosity for the similarity solutions in (A2).  Rearranging 
(A11) and substituting (A6) for $R_1$ gives
\begin{equation}
\nu_1 = \alpha c_s H \left(\frac{R}{R_c}\right)^{-\gamma} {\cal T}^{-\gamma/(2-\gamma)},
\end{equation}
which can be substituted into (A10) to solve for $\alpha$,
\begin{equation}
\alpha \approx \frac{2 R_c^2}{3(2-\gamma)} \frac{\dot{M}_{\ast}}{M_d} \left(\frac{R}{R_c}\right)^{\gamma} \frac{1}{c_s H} = \frac{\dot{M}_{\ast}}{3 \pi \Sigma_c} \left(\frac{R}{R_c}\right)^{\gamma} \frac{1}{c_s H}
\end{equation}
where the second equality has utilized equation (5) in \S 3.1.  This form in 
(A13) is identical to equation (10) in \S 5.1 \citep[see also][their equation 
22]{hartmann98}.  

\begin{figure}
\epsscale{0.5}
\plotone{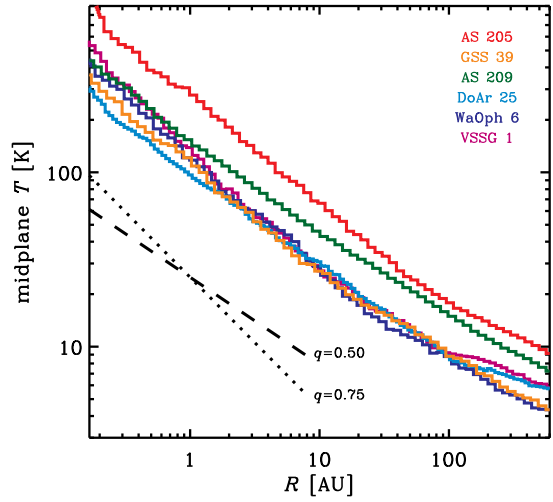}
\figcaption{Radial dust temperature profiles at the midplane for the sample 
disks with continuous density distributions, determined from radiative transfer 
calculations (see \S 3.3; the disks with central cavities are excluded for 
clarity).  The profiles are consistent with power-laws, $T \propto R^{-q}$ with 
$q \approx 0.5$-0.6, that flatten out at large radii ($R > R_c$).  Some 
representative power-laws are shown as dashed ($q = 0.50$) and dotted ($q = 
0.75$) lines in the lower left.  The midplane temperatures are set primarily by 
the vertical structure of the disk (see \S 4.1.4).  \label{Tmid}}
\end{figure}

\begin{figure}
\epsscale{0.45}
\plotone{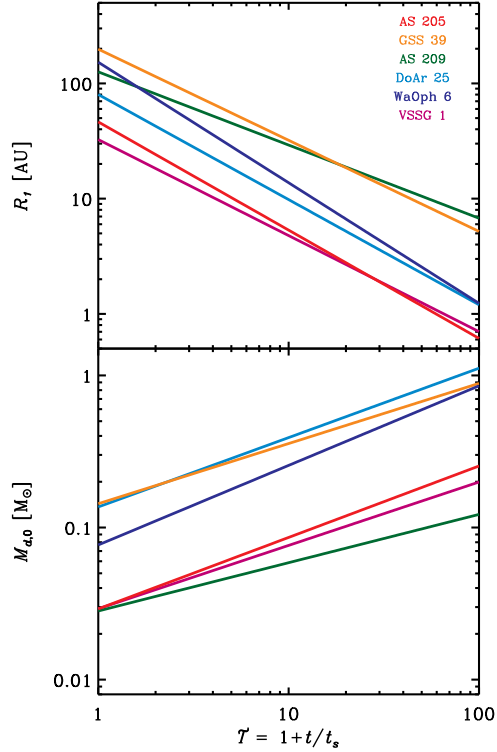}
\figcaption{Constraints on the initial conditions for a parametric viscous disk 
model as a function of the unknown value of ${\cal T}$ (the number of elapsed 
viscous timescales): ({\it top}) initial scaling radius from equation (7); and 
({\it bottom}) initial disk mass from equation (8).  Note that \{$R_1$, 
$M_{d,0}$\}, {\it do not vary with time} in these models.  Rather, these plots 
show what their appropriate values would be if ${\cal T}-1$ viscous timescales 
have elapsed since the start of the evolution process (see Table 
\ref{viscous_table}).  The disks with central cavities are excluded, because 
they have been clearly affected by additional evolutionary mechanisms.  
\label{accretion}}
\end{figure}

\begin{figure}
\epsscale{0.40}
\plotone{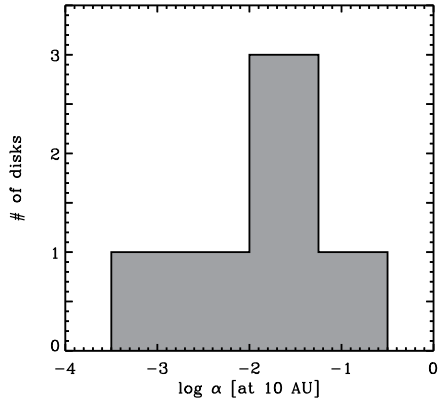}
\figcaption{Distribution of viscosity parameters, $\alpha$, inferred from the 
structure models and accretion rates for the sample disks, based on equation 
(10) (see Table \ref{viscous_table}).  While these values are appropriate for 
$R = 10$\,AU, $\alpha$ is only weakly dependent on radius (see the Appendix).  
\label{alphas}}
\end{figure}

\begin{figure}
\epsscale{0.5}
\plotone{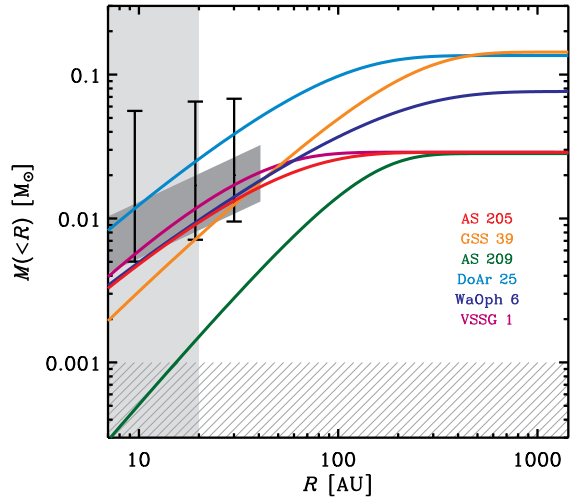}
\figcaption{The encircled mass profiles, or cumulative mass interior to $R$, 
for the sample disks with continuous density distributions.   Vertical error 
bars mark the range of acceptable $M$($<$$R$) values for Saturn, Uranus, and 
Neptune in the MMSN, reflecting the uncertainties of augmenting planet masses 
to solar composition (including those for planets interior to these radii).  
The dark gray region shows the standard $\Sigma \propto R^{-1.5}$ profiles for 
the MMSN with the range of normalizations adopted in the literature 
\citep{weidenschilling77,hayashi81}.  There is generally good agreement between 
the disk structure model results and the fossil record of the mass content for 
the primordial disk around the Sun.  The light shaded region marks the maximum 
resolution scale of the SMA observations, while the hatched region is an 
approximate representation of the 3\,$\sigma$ mass sensitivity of the data 
(assuming a typical $T \approx 10$\,K).  \label{Mcum}}
\end{figure}

Based on (A13) the value of $\alpha$ varies with radius proportional to 
$R^{\gamma}/c_s H$.  Following equation (3) in \S 3.1, the scale heights vary 
as $H \propto R^{1+\psi}$.  We can approximate the sound speed variation as 
$c_s \propto \sqrt{T}$, where $T$ is evaluated at the midplane.  Given the 
power-law temperature profiles found in \S 4.1.4 (see Fig.~\ref{Tmid}), $c_s 
\propto R^{-q/2}$.  Therefore, $\alpha$ has a power-law dependence on radius, 
$\alpha \propto R^z$ with $z = \gamma + 0.5q - \psi - 1$.  For the disks in 
this sample, $z$ is generally small ($< 0.1$) but can range from $-$0.4 (AS 
209) to 0.3 (WaOph 6).  At most, $\alpha$ varies by an order of magnitude 
across the disk, although typically it is constant within a factor of $\sim$2-3.

\clearpage

\begin{deluxetable}{lcccll}
\tablecolumns{6}
\tablewidth{0pc}
\tablecaption{SMA Observing Journal\label{obs_journal}}
\tablehead{
\colhead{Name} & \colhead{$\alpha$ [J2000]} & \colhead{$\delta$ [J2000]} & \colhead{Array} & \colhead{UT Date} & \colhead{Alt.~Name} \\
\colhead{(1)} & \colhead{(2)} & \colhead{(3)} & \colhead{(4)} & \colhead{(5)} & \colhead{(6)}}
\startdata
AS 205    & 16 11 31.35 & $-$18 38 26.0 & V & 2007 June 17 & V866 Sco, HBC 254 \\
           &             &               & V & 2007 May 26  \\
           &             &               & E & 2006 June 3  \\
           &             &               & C & 2006 May 12  \\
GSS 39     & 16 26 45.03 & $-$24 23 08.0 & V & 2008 April 5 & Elias 27 \\
           &             &               & C & 2006 May 14  \\
AS 209     & 16 49 15.29 & $-$14 22 08.8 & V & 2007 June 9 & V1121 Oph, HBC 270 \\
           &             &               & E & 2006 June 3  \\
           &             &               & C & 2006 May 12  \\
DoAr 25    & 16 26 23.69 & $-$24 43 14.1 & V & 2007 June 17 & YLW 34 \\
           &             &               & V & 2007 May 26  \\
           &             &               & C & 2005 June 12 \\
           &             &               & E & 2005 May 8   \\
WaOph 6    & 16 48 45.62 & $-$14 16 36.0 & V & 2007 June 9 & V2508 Oph, HBC 653 \\
           &             &               & C & 2005 June 12 \\
           &             &               & E & 2005 May 15  \\
SR 21      & 16 27 10.27 & $-$24 19 12.8 & V & 2007 June 10 & Elias 30 \\
           &             &               & S & 2008 August 29 & \\
VSSG 1     & 16 26 18.87 & $-$24 28 19.9 & C & 2008 May 13 & Elias 20 \\
           &             &               & V & 2008 April 5 \\
WSB 60     & 16 28 16.51 & $-$24 36 58.3 & C & 2008 May 13 & YLW 58 \\
           &             &               & V & 2008 April 3 \\
DoAr 44    & 16 31 33.46 & $-$24 27 37.4 & C & 2008 May 13 & Haro 1-16, ROXs 44 \\
           &             &               & V & 2008 April 3 
\enddata
\tablecomments{Col.~(1): Disk name.  Cols.~(2) \& (3): Centroid coordinates, 
determined as described in the text (\S 2).  Col.~(4): Array configuration; V = 
very extended (68-509\,m baselines), E = extended (28-226\,m baselines), C = 
compact (16-70\,m baselines), and S = sub-compact (6-70\,m baselines).  
Col.~(5): UT date of observation.  Observations prior to 2007 were described by 
\citet{aw07}, and the archival SR 21 V data were originally presented by 
\citet{brown09}.  Col.~(6): Common alternative identifications.}
\end{deluxetable}

\clearpage

\begin{deluxetable}{lcccccccc}
\tablecolumns{9}
\tablewidth{0pc}
\tablecaption{Continuum and CO $J$=3$-$2 Synthesized Map Properties\label{data_table}}
\tablehead{
\colhead{Disk} & \multicolumn{4}{c}{continuum} & \colhead{} & \multicolumn{3}{c}{CO $J$=3$-$2} \\
\cline{2-5} \cline{7-9} 
\colhead{} & \colhead{$\lambda_{{\rm eff}}$} & \colhead{$F_{\nu}$} & \colhead{$\theta_b$} & \colhead{PA$_b$}  & \colhead{} & \colhead{rms} & \colhead{$\theta_b$} & \colhead{PA$_b$} \\
\colhead{} & \colhead{[$\mu$m]} & \colhead{[mJy]} & \colhead{[\arcsec]} & \colhead{[\degr]} & \colhead{} & \colhead{[Jy]} & \colhead{[\arcsec]} & \colhead{[\degr]} \\
\colhead{(1)} & \colhead{(2)} & \colhead{(3)} & \colhead{(4)} & \colhead{(5)} & \colhead{} & \colhead{(6)} & \colhead{(7)} & \colhead{(8)}}
\startdata
AS 205   & 859 & $960\pm7$ & $0.46\times0.37$ & 22 & & 0.25 & $2.19\times1.60$ & 176 \\
GSS 39   & 882 & $663\pm3$ & $0.63\times0.54$ & 26 & & 0.50 & $3.26\times1.80$ & 168 \\
AS 209   & 859 & $577\pm3$ & $0.56\times0.45$ & 36 & & 0.25 & $2.38\times1.73$ & 176 \\
DoAr 25  & 865 & $563\pm3$ & $0.45\times0.34$ & 15 & & 0.22 & $1.80\times1.43$ & 178 \\
WaOph 6  & 869 & $405\pm3$ & $0.49\times0.38$ & 37 & & 0.20 & $1.61\times1.44$ & 177 \\
SR 21    & 865 & $371\pm2$ & $0.50\times0.36$ & 17 & & 0.13 & $3.07\times2.03$ & 3   \\
VSSG 1   & 882 & $275\pm2$ & $0.46\times0.30$ & 31 & & 0.22 & $2.36\times1.32$ & 47  \\
WSB 60   & 883 & $255\pm3$ & $0.58\times0.38$ & 25 & & 0.38 & $2.96\times2.42$ & 31  \\
DoAr 44  & 883 & $229\pm3$ & $0.57\times0.35$ & 23 & & 0.26 & $2.29\times1.35$ & 47  
\enddata
\tablecomments{Col.~(1): Disk name.  Col.~(2): Effective wavelength of the 
combined continuum data.  Col.~(3): Integrated continuum flux density and rms 
noise in the map (does not include the $\sim$10\%\ flux calibration 
uncertainty).  Cols.~(4) \& (5): FWHM dimensions and position angle (measured 
east of north) of the synthesized beam for the continuum maps shown in Figure 
\ref{images}.  Note that the SR 21 images were generated for a different 
visibility weighting (robust = 0.3, compared to 0.7 for the rest of the 
sample).  Col.~(6): The rms noise level per beam for an individual channel 
in the synthesized channel maps.  For AS 205 and AS 209, the channel width is 
0.18\,km s$^{-1}$, for SR 21 it is 0.44\,km s$^{-1}$, and for the others it is 
0.70\,km s$^{-1}$.  Cols.~(7) \& (8): FWHM dimensions and position angle of the 
synthesized beam for the CO channel maps.}
\end{deluxetable}

\begin{deluxetable}{lcccccccc}
\tablecolumns{10}
\tablewidth{0pt}
\tablecaption{Stellar Properties\label{stars_table}}
\tablehead{
\colhead{Name} & \colhead{SpT} & \colhead{$A_V$} & \colhead{$T_{{\rm eff}}$} & \colhead{$R_{\ast}$} & \colhead{$L_{\ast}$} & \colhead{$M_{\ast}$} & \colhead{age} & \colhead{ref} \\
\colhead{} & \colhead{} & \colhead{[mag]} & \colhead{[K]} & \colhead{[R$_{\odot}$]} & \colhead{[L$_{\odot}$]} & \colhead{[M$_{\odot}$]} & \colhead{[Myr]} & \colhead{} \\
\colhead{(1)} & \colhead{(2)} & \colhead{(3)} & \colhead{(4)} & \colhead{(5)} & \colhead{(6)} & \colhead{(7)} & \colhead{(8)} & \colhead{(9)}}
\startdata
AS 205   & K5 & 2.9  & 4250 & 3.7 & 4.0 & 1.0 & 0.5 & 1 \\
GSS 39   & M0 & 15   & 3850 & 2.3 & 1.0 & 0.6 & 1.0 & 2 \\
AS 209   & K5 & 0.9  & 4250 & 2.3 & 1.5 & 0.9 & 1.6 & 3 \\
DoAr 25  & K5 & 2.7  & 4250 & 1.7 & 0.8 & 1.0 & 3.8 & 4 \\
WaOph 6  & K6 & 3.6  & 4205 & 3.2 & 2.9 & 0.9 & 0.7 & 5 \\
SR 21    & G3 & 6.3  & 5800 & 3.3 & 11  & 2.0 & 4.7 & 1 \\
VSSG 1   & M0 & 14   & 3850 & 3.1 & 1.9 & 0.6 & 0.7 & 6 \\
WSB 60   & M4 & 3.5  & 3370 & 1.3 & 0.2 & 0.3 & 3.0 & 4 \\
DoAr 44  & K3 & 2.3  & 4730 & 1.7 & 1.3 & 1.4 & 7.1 & 7 
\enddata
\tablecomments{Col.~(1): Disk name.  Col.~(2): Spectral type.  Col.~(3): Visual 
extinction.  Col.~(4): Effective temperature.  Col.~(5): Stellar radius.  
Col.~(6): Stellar luminosity.  Col.~(7) and (8): Stellar mass and age estimated 
from the \citet{siess00} pre-main-sequence models.  Col.~(9): Literature 
references for SpT and $A_V$: [1] - \citet{prato03}, [2] - \citet{luhman99}, 
[3] - \citet{herbig88}, [4] - \citet{wilking05}, [5] - \citet{eisner05}, [6] - 
\citet{natta06}, [7] - \citet{bouvier92}.}
\end{deluxetable}

\begin{deluxetable}{lccccc|ccc|cc}
\tablecolumns{11}
\tablewidth{0pt}
\tablecaption{Disk Structure Model Parameters: Continuous Cases\label{structure_table}}
\tablehead{
\colhead{Name} & \colhead{$M_d$} & \colhead{$\gamma$} & \colhead{$R_c$} & \colhead{$H_{100}$} & \colhead{$\psi$} & \colhead{$R_{{\rm in}}$} & \colhead{$i$} & \colhead{PA} & \colhead{$\tilde{\chi}^2_{\rm vis}$} & \colhead{$\tilde{\chi}^2_{\rm sed}$} \\
\colhead{} & \colhead{[M$_{\odot}$]} & \colhead{} & \colhead{[AU]} & \colhead{[AU]} & \colhead{} & \colhead{[AU]} & \colhead{[\degr]} & \colhead{[\degr]} & \colhead{} & \colhead{} \\
\colhead{(1)} & \colhead{(2)} & \colhead{(3)} & \colhead{(4)} & \colhead{(5)} & \colhead{(6)} & \colhead{(7)} & \colhead{(8)} & \colhead{(9)} & \colhead{(10)} & \colhead{(11)}}
\startdata
AS 205   & 0.029 & 0.9 & 46  & 19.6 & 0.11 & 0.14 & 25 & 165 & 2.1 & 3.7 \\  
GSS 39   & 0.143 & 0.7 & 198 & 7.3  & 0.08 & 0.07 & 60 & 110 & 1.9 & 32  \\  
AS 209   & 0.028 & 0.4 & 126 & 13.3 & 0.10 & 0.09 & 38 & 86  & 1.7 & 2.4 \\  
DoAr 25  & 0.136 & 0.9 & 80  & 6.7  & 0.15 & 0.06 & 59 & 112 & 1.9 & 9.2 \\  
WaOph 6  & 0.077 & 1.0 & 153 & 4.4  & 0.06 & 0.12 & 39 & 171 & 1.8 & 1.8 \\  
VSSG 1   & 0.029 & 0.8 & 33  & 9.7  & 0.08 & 0.10 & 53 & 165 & 1.8 & 12  \\  
\enddata
\tablecomments{Col.~(1): Disk name.  Col.~(2): Disk mass assuming a 100:1
gas-to-dust mass ratio.  Col.~(3): Radial surface density gradient (eqn.~4).
Col.~(4): Characteristic radius (eqns.~3-4).  Col.~(5): Scale height at
100\,AU (eqn.~2).  Col.~(6): Radial scale height gradient (eqn.~3).  Col.~(7):
Fixed inner disk radius (eqn.~6).  Col.~(8): Fixed disk inclination (see \S 
3.2).  Col.~(9): Fixed major axis position angle (see \S 3.2).  Col.~(10): 
Reduced $\chi^2$ statistic ($\chi^2$ divided by the number of degrees of 
freedom) comparing the model fit with the continuum visibilities alone.  
Col.~(11): Same as Col.~(10), but for the SED alone.}
\end{deluxetable}

\begin{deluxetable}{lccccccc|cc|cc}
\tablecolumns{12}
\tablewidth{0pt}
\tablecaption{Disk Structure Model Parameters: Central Cavity Cases\label{trans_table}}
\tablehead{
\colhead{Name} & \colhead{$M_d$} & \colhead{$\gamma$} & \colhead{$R_c$} & \colhead{$H_{100}$} & \colhead{$\psi$} & \colhead{$R_{\rm cav}$} & \colhead{$\delta_{\rm cav}$} & \colhead{$i$} & \colhead{PA} & \colhead{$\tilde{\chi}^2_{\rm vis}$} & \colhead{$\tilde{\chi}^2_{\rm sed}$} \\
\colhead{} & \colhead{[M$_{\odot}$]} & \colhead{} & \colhead{[AU]} & \colhead{[AU]} & \colhead{} & \colhead{[AU]} & \colhead{} & \colhead{[\degr]} & \colhead{[\degr]} & \colhead{} & \colhead{} \\
\colhead{(1)} & \colhead{(2)} & \colhead{(3)} & \colhead{(4)} & \colhead{(5)} & \colhead{(6)} & \colhead{(7)} & \colhead{(8)} & \colhead{(9)} & \colhead{(10)} & \colhead{(11)} & \colhead{(12)}}
\startdata
SR 21    & 0.005 & 0.9 & 17  & 7.7  & 0.26 & 37 & 0.005  & 22 & 110 & 1.7 & 7.2 \\
WSB 60   & 0.021 & 0.8 & 31  & 11.0 & 0.13 & 20 & 0.01   & 25 & 117 & 1.8 & 3.0 \\
DoAr 44  & 0.017 & 1.0 & 80  & 3.5  & 0.04 & 33 & 0.0001 & 45 & 75  & 1.8 & \nodata \\
\enddata
\tablecomments{Cols.~(1-6): Same as for Table \ref{structure_table}.  Col.~(7): 
The cavity radius, marking the outer edge of the diminished inner disk 
densities (see \S 3.1 and \S 4.1.5).  Col.~(8): The density reduction scaling 
factor inside the radius $R_{\rm cav}$ (see \S 3.1 and \S 4.1.5).  
Cols.~(9-12): Same as for Table \ref{structure_table} cols.~(8-11).}
\end{deluxetable}

\clearpage

\begin{deluxetable}{lcccl|cccc}
\tablecolumns{9}
\tablewidth{0pt}
\tablecaption{Viscous Disk Properties\label{viscous_table}}
\tablehead{
\colhead{Name} & \colhead{$\dot{M}_{\ast}$} & \colhead{ref} & \colhead{$R_t$} & \colhead{$\alpha$} & \multicolumn{2}{c}{$R_1$} & \multicolumn{2}{c}{$M_{d,0}$} \\ \colhead{} & \colhead{[M$_{\odot}$ yr$^{-1}$]} & \colhead{} & \colhead{[AU]} & \colhead{} & \multicolumn{2}{c}{[AU]} & \multicolumn{2}{c}{[M$_{\odot}$]} \\ \colhead{(1)} & \colhead{(2)} & \colhead{(3)} & \colhead{(4)} & \colhead{(5)} & \colhead{(6)} & \colhead{(7)} & \colhead{(8)} & \colhead{(9)}}
\startdata
AS 205   & $8\times10^{-8}$  & 1       & 23 & 0.005   & 24  & 5  & 0.040 & 0.090 \\
GSS 39   & $7\times10^{-8}$  & 2       & 95 & 0.03    & 115 & 30 & 0.189 & 0.370 \\
AS 209   & $9\times10^{-8}$  & 3       & 61 & 0.08    & 81  & 27 & 0.035 & 0.061 \\
DoAr 25  & $3\times10^{-9}$  & 4       & 39 & 0.0005  & 43  & 9  & 0.187 & 0.407 \\
WaOph 6  & $1\times10^{-7}$  & 5       & 78 & 0.05    & 74  & 12 & 0.110 & 0.269 \\
VSSG 1   & $1\times10^{-7}$  & 2       & 16 & 0.02    & 18  & 4  & 0.039 & 0.079 \\
\hline
SR 21    & $< 2\times10^{-9}$ & 2      & 9  & \nodata & 9   & 2  & 0.007 & 0.015 \\
WSB 60   & $2\times10^{-9}$  & 2       & 15 & \nodata & 18  & 5  & 0.027 & 0.054 \\
DoAr 44  & \nodata           & \nodata & 40 & \nodata & 40  & 7  & 0.023 & 0.055 \\
\enddata
\tablecomments{Col.~(1): Disk name.  Col.~(2): Mass accretion rate.  Col.~(3): 
Literature references for $\dot{M}_{\ast}$: [1] - \citet{prato03}, [2] - 
\citet{natta06}, [3] - \citet{johnskrull00}, [4] - \citet{luhman99}, [5] - 
\citet{eisner05}.  Col.~(4): Radius marking where the mass flow changes 
direction, based on equation (9).  Col.~(5): Viscosity parameter, based on 
equation (10).  Col.~(6): Initial radius encircling $\sim$60\%\ of the disk 
mass if 1 viscous timescale has elapsed (${\cal T} = 2$), based on equation (7) 
(see Fig.~\ref{accretion}).  Col.~(7): Same as Col.~(6), but for the case where 
10 viscous timescales have elapsed (${\cal T} = 11$).  Col.~(8): Initial disk 
mass if 1 viscous timescale has elapsed (${\cal T} = 2$), based on equation (8) 
(see Fig.~\ref{accretion}).  Col.~(9): Same as Col.~(8), but for the case where 
10 viscous timescales have elapsed (${\cal T} = 11$).}
\end{deluxetable}

\clearpage

\clearpage

\clearpage

\clearpage

\clearpage

\clearpage

\clearpage

\clearpage

\clearpage

\end{document}